\documentclass[11pt]{article}
\usepackage{natbib}
\usepackage{epsfig}
\usepackage{graphicx}
\usepackage{amsthm,amssymb,amsmath}
\usepackage{amsfonts}
\usepackage{verbatim,enumerate}
\usepackage{verbatim,multicol,color}
\usepackage{latexsym}
\usepackage{float}
\usepackage{indentfirst}       
\usepackage{booktabs}
\usepackage{subfigure}

\DeclareGraphicsExtensions{.jpg,.pdf,.eps,.png}

\def\JRSSB{{\it Journal of the Royal Statistical Society, Series B}}

\def\JRSSB{{\it Journal of the Royal Statistical Society, Series B}}

\def\JRSSB{{\it Journal of the Royal Statistical Society, Series B}}

\def\boxit#1{\vbox{\hrule\hbox{\vrule\kern6pt
          \vbox{\kern6pt#1\kern6pt}\kern6pt\vrule}\hrule}}

\def\bse{\begin{eqnarray*}}
\def\ese{\end{eqnarray*}}
\def\be{\begin{eqnarray}}
\def\ee{\end{eqnarray}}
\def\bq{\begin{equation}}
\def\eq{\end{equation}}
\def\bse{\begin{eqnarray*}}
\def\ese{\end{eqnarray*}}

\newtheorem{Definition}{Definition}

\newtheorem{prop}{Proposition}

\newtheorem{cor}{Corollary}

\def\part{\partial}

\def\RR{\mathbb{R}}

\newcommand{\bT}{{\bf T}}

\newcommand{\bU}{{\bf U}}

\newcommand{\bX}{{\bf X}}
\newcommand{\bx}{{\bf x}}

\newcommand{\bZ}{{\bf Z}}
\newcommand{\ba}{{\bf a}}

\newcommand{\bD}{{\bf D}}

\newcommand{\bomega}{\mbox{\protect\boldmath $\omega$}}
\newcommand{\btheta}{\mbox{\protect\boldmath $\theta$}}

\newcommand{\bbeta}{\mbox{\boldmath $\beta$}}

\newcommand{\bgamma}{\mbox{\boldmath $\gamma$}}
\newcommand{\bGamma}{\mbox{\boldmath $\Gamma$}}
\newcommand{\bDelta}{\mbox{\boldmath $\Delta$}}

\newcommand{\bLambda}{\mbox{\boldmath $\Lambda$}}

\newcommand{\bmu}{\mbox{\protect\boldmath $\mu$}}
\newcommand{\bfeta}{\mbox{\protect\boldmath $\eta$}}
\newcommand{\bOmega}{\mbox{\protect\boldmath $\Omega$}}

\newcommand{\bSigma}{\mbox{\protect\boldmath $\Sigma$}}

\newcommand{\bI}{\mbox{\protect\boldmath $I$}}
\newcommand{\bepsilon}{\mbox{\protect\boldmath $\epsilon$}}
\newcommand{\bV}{\mbox{\protect\boldmath $V$}}

\newcommand{\by}{{\bf y}}
\newcommand{\bY}{{\bf Y}}

\begin{document}
\thispagestyle{empty}
\baselineskip=28pt
\vskip 5mm
\begin{center} {\LARGE{\bf Multivariate Log-Skewed Distributions with
normal kernel and their Applications
}}
\end{center}

\baselineskip=12pt
\vskip 10mm

\begin{center}\large
Marina M. de Queiroz \footnote{Departamento de Estat\'{\i}stica,
 Universidade Federal de Minas Gerais,
 31270-901 - Belo Horizonte - MG, Brazil. E-mail: marinamunizdequeiroz@gmail.com.
 }, Rosangela H. Loschi \footnote{Departamento de Estat\'{\i}stica,
 Universidade Federal de Minas Gerais,
 31270-901 - Belo Horizonte - MG, Brazil. E-mail: loschi@est.ufmg.br}, Roger W. C. Silva \footnote{
\baselineskip=10pt Departamento de Estat\'{\i}stica,
 Universidade Federal de Minas Gerais,
 31270-901 - Belo Horizonte - MG, Brazil. E-mail: rogerwcs@est.ufmg.br (corresponding author).
 }
\\ 
Departamento de Estat\'{\i}stica,
Universidade Federal de Minas Gerais.
\baselineskip=15pt
\vskip 10mm
\centerline{\today}
\vskip 10mm

\end{center}

\begin{abstract}
We  introduce two classes of  multivariate log skewed
distributions with normal kernel: the log canonical fundamental
skew-normal (log-CFUSN) and the log unified skew-normal (log-SUN).
We also discuss some properties of the log-CFUSN family of distributions.
These new classes of log-skewed distributions include
the log-normal and multivariate log-skew normal families as
particular cases. We discuss some issues related to Bayesian
inference in the log-CFUSN family of distributions, mainly we focus on how to model
the prior uncertainty about the skewing parameter. Based on the
stochastic representation of the log-CFUSN family, we propose
a data augmentation strategy for sampling from the posterior
distributions. This proposed family is used to analyze the US
national monthly precipitation data. We conclude that a high
dimensional skewing function lead to a better model fit.
\newline


\end{abstract}



\section{Introduction}
\label{sec.intro}


The construction of new parametric distributions has received considerable attention in recent years.
This growing interest is motivated by datasets that often present strong skewness, heavy tails,
bimodality and some other characteristics that are not well fitted by the usual distributions,
such as the normal, Student-$t$, log-normal, exponential  and many others. The main goal
is to build more flexible parametric distributions with additional parameters
allowing to control such characteristics. 
If compared to finite  mixtures of distributions \citep[see][for instance]{LiLeYe07, CaBoPe08}
or nonparametric methods \citep[for recent surveys on Bayesian nonparametric see][]{MuQu04, Wa05, Dey98},
one advantage of this approach is that, in general, more parsimonious models are  obtained and, as a consequence,
the inference process tends to become simpler.

It is not feasible to mention all developments
in this area in recent years. \cite{ArBe02}, \cite{Ge04} and \cite{Az05} review several recent
works in the area and  are important sources of a detailed discussion of such distributions properties.
Further advances in the area can be found in   \cite{GeLo05}, \cite{ArAz06}, \cite{ArBrGe06},
\cite{ArGoSa09}, \cite{ElGoQu09}, \cite{ArCoGo10}, \cite{MaGe10}, \cite{GoElSaBo11},
\cite{BoGoRi11}, \cite{RoLoAr13} and many others.

The seminal paper by \cite{Az85} is one of the main references
in this topic and has inspired many other works.
\cite{Az85} introduced the so called skew-normal (SN) family of distributions
which  probability density function (pdf) is  
\begin{equation}
\label{skewnormal}
f(z \mid \mu,\omega,\alpha)=\frac{2}{\omega}\phi\left(\frac{z-\mu}{\omega}\right)
\Phi\left(\alpha\left(\frac{z-\mu}{\omega}\right)\right),\,\,z \in \mathbb{R},
\end{equation}
where $\mu\in\mathbb{R}$ and  $\omega\in\mathbb{R}^+$ are the location and scale parameters, respectively,  $\alpha \in\mathbb{R}$ is
the skewness  parameter and $\phi$ and $\Phi$ denote, respectively, the pdf and the
cumulative  distribution function (cdf) of the $N(0,1)$.
The family in (\ref{skewnormal}) extends the normal one by introducing an extra parameter
to control the asymmetry of the distribution and has the normal family as a particular subclass
whenever $\alpha$ equals zero. It also preserves some nice properties of the normal family. Another
 extension  of the univariate distribution in  (\ref{skewnormal}) recently appeared in \cite{MaBo14} which introduced the so called skew-normal alpha-power distibution.
The multivariate analog of the SN distribution was introduced by \cite{AzDa96}.

In a more general setting, \cite{GeLo05} introduced the class of generalized
multivariate skew elliptical (GSE) distributions which pdf is
\begin{equation}\label{fundamental}
f(\mathbf{z}|Q)=2f_k(\mathbf{z})Q(\mathbf{z}),\,\,\mathbf{z}\in\RR^k,
\end{equation}
where $f_k$ is the pdf of a $k$-dimensional elliptical distribution and $Q$
is a skewing function satisfying $Q(-\mathbf{z})=1-Q(\mathbf{z})$, for all $\mathbf{z}\in \RR^k$.
Many of the SN distribution properties also follow to any distribution
in this class. Particularly,   \cite{GeLo05} prove that
distributions  of quadratic forms  in the GSE family do not  depend on the
skewing function $Q$. Some other properties of the  GSE family, such as the joint moment generating functions of  linear transformations and quadratic forms of $\bZ$ and the conditions for their independence, 
can be found in \cite{HuSuGu13}.
It should be also mentioned that the multivariate SN families of distributions
defined by \cite{AzDa96} and \cite{AzCa99} 
and  the family  of skew-spherical (elliptical) distributions defined in \cite{BrDe01}
are subclasses  of (\ref{fundamental}).   

\cite{AzDa96}'s  family of distributions is also a subclass of the fundamental SN (FUSN)
class of distributions defined by \cite{ArGe05}.  A vector  $Z^*$ has a $n$-variate canonical
fundamental skew-normal (CFUSN) distribution with an $n \times m$
skewness matrix $\bDelta$, which will be denoted by $Z^*\sim CFUSN_{n,m}(\bDelta)$, if its density is given by
\begin{equation}
\label{CFUSN}
f_{\bZ^*}(\mathbf{z})=2^m\phi_n(\mathbf{z})\Phi_m(\bDelta'\mathbf{z}|{\bf{I}}_m-\bDelta' \bDelta),\,\,\,\,\mathbf{z}\in \RR^n,
\end{equation}
where $\bDelta$ is such that $||\bDelta \mathbf{a}|| < 1$, for all unitary vectors $\mathbf{a}\in \RR^m$, and $||^{.}||$
denotes euclidean norm. Along  this paper, we  denote by $\phi_n({\mathbf y}\mid{\mathbf
\bmu},{\mathbf \Sigma})$ the p.d.f. associated with the multivariate
$N_n({\mathbf \bmu},{\mathbf \Sigma})$ distribution, and by
$\Phi_n({\mathbf y}\mid{\mathbf \bmu},{\mathbf \Sigma})$ the
corresponding cumulative distribution function (c.d.f.). If $\mathbf{\bmu}={\mathbf 0}$ (respectively
$\mathbf{\bmu}={\mathbf 0}$ and ${\mathbf \Sigma} ={\bf{I}}_n$) these
functions will be denoted by $\phi_n({\mathbf y}\mid{\mathbf
\Sigma})$ and $\Phi_n({\mathbf y}\mid{\mathbf \Sigma})$
(respectively $\phi_n({\mathbf y})$ and $\Phi_n({\mathbf y})$).
For simplicity, $\phi({y})$ and $\Phi({y})$ will
be used in the univariate case.

Several classes of SN distributions were defined in the literature.
An unification of these families is proposed by \cite{ArAz06} which define the unified skew-normal  family of distribution,
the so-called SUN family.
A random vector  $Z^* \sim SUN_{n,m}(\bfeta,\bgamma, \bar{\bomega}, \bOmega^*)$ if its pdf is 
\begin{equation}
\label{DeSUN}
f_{\bZ^*}(\mathbf{z})=\phi_n(\mathbf{z}- \bfeta\mid \bOmega)
\frac{\Phi_m(\bgamma + \bDelta'\bar{\bOmega}^{-1}\bomega^{-1}(\mathbf{z}- \bfeta)|\bGamma - \bDelta'\bar{\bOmega}^{-1} \bDelta)}
{\Phi_m(\bgamma \mid \bGamma)},\,\,\,\,\mathbf{z}\in\RR^n,
\end{equation}
where the vectors  $\bfeta \in \RR^n$ and $\bgamma \in \RR^m$,
$\bar{\bomega}$ is the vector of the diagonal elements of $\omega$, $\omega$ is a diagonal matrix formed by the standard deviations of
$\bOmega = \bomega \bar{\bOmega} \bomega$, $\bar{\bOmega}$,  $\bGamma$ and $\bDelta$ are, respectively,  $n \times n$,  $m \times m$
and $n \times m$  matrices such that
$$
\bOmega^* = \left(
\begin{array}{cc}
\bGamma & \bDelta' \\
\bDelta & \bar{\bOmega} \\
\end{array}\right)
$$
is a correlation matrix. For another unification of multivariate skewed distributions see \cite{AbTo13}.

In limit cases,  some of these distributions concentrate their
probability mass in positive (or negative) values.  The half-normal distribution,
for instance, is obtaind from  (\ref{skewnormal}) by assuming $\alpha$
equal to infinite. Because of this,  such family of distributions has also been
considered to  model data with positive support, such as income, precipitation,
pollutants concentration and so on. However, such limit distributions are not flexible enough to accommodate the diversity of shapes of positive (or negative) data. In the univariate context, Gamma, exponential
and log-normal distributions are commonly used to model non-negative random variables.
Less conventional analysis can be done using the log-SN and log-Skew-$t$ 
introduced by \cite{AzCaKo03} or the log-power-normal distribution introduced by \cite{MaBo12}. 

In the multivariate context, however, distributions with positive support
are usually intractable,  with the exception of the multivariate log-normal
distribution. With the above problem in mind, \cite{MaGe10}  built the
multivariate log-skew elliptical family of distributions  as follows. Denote
by $El_n(\mathbf{\mu},\mathbf{\Sigma},g^{(n)})$ the family of $n$-dimensional elliptical
distributions (with existing pdf) with generating function
$g^{(n)}(u), u\geq 0$, defining a $n$-dimensional spherical density, a location column vector $\mathbf{\mu} \in \mathbb{R}^n$,
and a $n\,$x$\,n$ positive definite
dispersion matrix $\mathbf{\Sigma}$. If $\mathbf{X}\sim El_n(\mathbf{\mu},\mathbf{\Sigma},g^{(n)})$, then its pdf is $
f_{n}(\mathbf{x}; \mathbf{\mu},\mathbf{\Sigma}, g^{(n)})=|\mathbf{\Sigma}|^{-\frac{1}{2}}g^{(n)}(Q_{\mathbf x}^{\mathbf{\mu},
\mathbf{\Sigma}})$, where $Q_{\mathbf x}^{\mathbf{\mu},\mathbf{\Sigma}}=(\mathbf{x}-\mathbf{\mu})'\mathbf{\Sigma}^{-1}
(\mathbf{x}-\mathbf{\mu})$, $\mathbf{x}\in\mathbb{R}^n$ \citep{Fang90}. Consider the class of skew elliptical distributions
with pdf given by 
\begin{equation}\label{skewelliptical}
f_{SEl_{n}}(\mathbf{x})=2f_{n}(\mathbf{x}; \mathbf{\mu},\mathbf{\Sigma}, g^{(n)})F(\mathbf{\alpha'}
\mathbf{\omega^{-1}}(\mathbf{x}-\mathbf{\mu});g_{Q_{\mathbf{x}}^{\mathbf{\mu},\mathbf{\Sigma}}}),\,\,  \mathbf{x}  \in \mathbb{R}^{n},
\end{equation}
where $\mathbf{\alpha} \in \mathbb{R}^n$ is a shape parameter, $\mathbf{\omega}=diag(\mathbf{\Sigma})^{1/2}$ is a
$n\,$x$\,n$ scale matrix, $f_{n}(\mathbf{x}; \mathbf{\mu},\mathbf{\Sigma}, g^{(n)})$ is the pdf of a $n$-dimensional
random vector of $El_{n}(\mathbf{\mu}, \mathbf{\Sigma}, g^{(n)})$ and $F(u;g_{Q_{\mathbf x}^{\mathbf{\mu},\mathbf{\Sigma}}})$
is the cdf of the $El_{1}(0,1,g_{Q_{\mathbf x}^{\mathbf{\mu},\mathbf{\Sigma}}})$ with generating function
$g_{Q_{\mathbf x}^{\mathbf{\mu},\mathbf{\Sigma}}}(u)=g^{(n+1)}(u+ Q_{\mathbf x}^{\mathbf{\mu},\mathbf{\Sigma}})/g^{(n)}
(Q_{\mathbf x}^{\mu,\Sigma})$. The distribution in (\ref{skewelliptical})
is denoted by $SEl_n(\mathbf{\mu},\mathbf{\Sigma}, \mathbf{\alpha},g^{(n+1)})$.
Consider the transformation
${\exp(\mathbf{X})}=(\exp(X_1),\dots,\exp(X_n))$, where $\mathbf{X}\sim SEl_n(\mathbf{\mu},\mathbf{\Sigma},
\mathbf{\alpha},g^{(n+1)})$. Then, $\mathbf{X}$ has log-skew elliptical distribution denoted by $\mathbf{X} \sim LSEl_n(\mathbf{\mu},\mathbf{\Sigma},
\mathbf{\alpha},g^{(n+1)})$ with pdf
\begin{equation}\label{logskewelliptical}
f_{LSEl_{n}}(\mathbf{x})=2\left(\prod_{i=1}^{n}{x_{i}}\right)^{-1}f_{n}(\ln(\mathbf{x}); \mathbf{\mu},\mathbf{\Sigma}, g^{(n)})F(\mathbf{\alpha'}
\mathbf{\omega^{-1}}(\ln(\mathbf{x})-\mathbf{\mu});g_{Q_{\mathbf{x}}^{\mathbf{\mu},\mathbf{\Sigma}}}), \  \mathbf{x}>0.
\end{equation}
%

It is immediate that the multivariate skew-normal \citep{AzDa96} and skew-t \citep{AzCa03}
distributions are special cases of (\ref{skewelliptical}).  Consequently, the log-skewed
class of distributions in (\ref{logskewelliptical}) introduced by \cite{MaGe10}  also defines
particular classes of multivariate log-SN and log-skew-$t$ distributions and has, as a special
case, the multivariate log-normal family of distributions.

Our main motivation to introduce new classes of multivariate log-skewed distribution  are 
some results that recently appeared in a  paper by \cite{SaLoAr13}. That paper focused on the
parameter interpretation in the mixed logistic regression models which  is done through the
so called odds ratio as in the usual logistic regression model. However, by considering
the random effects,   the odds ratio to compare two individuals in two  different clusters
becomes a random variable ($OR$) that depends on the  random effects related to the two clusters
under comparison \citep{LaPeJoEn00}.  Because of this, \cite{LaPeJoEn00}
propose to interpret the odds ratio in terms of the median of its distribution
in order to quantify appropriately the heterogeneity among the different clusters.
If the random effects are independent and identically distributed (iid) with
$SN(\xi,\sigma^2,\lambda)$ then \cite{SaLoAr13} prove that the odds ratio has distribution with pdf
given by
\begin{equation}
f_{OR|\bbeta,\btheta, \bx}(r) = \frac{4}{r}\phi(\ln{r}|\kappa_{12},2\sigma^2) 
\times\Phi_2\left(\frac{\delta\ln{r}}{2\sigma}\bepsilon|
\frac{\delta\kappa_{12}}{2\sigma}\bepsilon,
 {\bf{I}}_2 - \frac{\delta^{2}}{2}\bepsilon\bepsilon' \right), \;\;r\in\RR_{+},
\label{EqLNA11}
\end{equation}
where $\kappa_{12}=(\bx'_{i_1j_1}-\bx'_{i_2j_2})\bbeta$,
$\bepsilon = (1, -1)'$ and $\delta = \lambda(1+ \lambda^2)^{-0.5}$.
Similar distributions were also obtained under independent skew-normally distributed  random effects.
The univariate log-skewed  distribution in
(\ref{EqLNA11}) does not belong to the class of distributions  defined by \cite{MaGe10},
nor  to that introduced by \cite{AzCaKo03}. Moreover, only its median was obtained by  \cite{SaLoAr13}
but no other property of it was studied.

In this paper, we introduce the multivariate log-CFUSN  and log-SUN  family of
distributions.  We explore their relationship  and study some properties of the log-CFUSN
family of distributions. Such classes of distributions have
as subclasses the multivariate log-skew-normal family introduced by \cite{MaGe10},
the log-SN family by \cite{AzCaKo03} and the family of distributions given in (\ref{EqLNA11}).
We also discuss some issues related to Bayesian inference in this
family. To illustrate its use we analyze the USA monthly precipitation data recorded from 1895 to 2007,
that is available at the National Climatic Data Center (NCDC).

This paper is organized as follows. In Section \ref{Sec2} we define the log-CFUSN and the log-SUN families of  distributions
and establish some of the probabilistic properties of the log-CFUSN  family of distributions.
Bayesian inference in the log-CFUSN family is discussed in Section \ref{Sec3}.
In Section \ref{SecCS} we present some data analysis using the proposed
log-CFUSN family of distributions. Finally, Section \ref{SecCo}
finishes the paper with a discussion and our main conclusions.

\section{Log-SUN and Log-CFUSN families of distributions}
\label{Sec2}

Under the normal theory, the log-normal family of distributions is obtained
assuming the logarithimic transformation. If a random variable $Y$ is log-normally distributed
it follows that the log transformation of it, that is, $X=\ln Y$, has a normal
distribution. Following this idea, in this section, we formally define the log-canonical-fundamental-skew-normal (log-CFUSN)
and the log-unified-skew-normal (log-SUN) families of  distributions
and explore some properties of the log-CFUSN such as conditional and marginal distributions, mixed moments and
stochastic representations.

Let   $\bZ^{*}=(Z^{*}_1, \dots, Z^{*}_n)'$ be an $n \times 1$ random vector and consider the transformations
${\exp({\mathbf{Z}}^*)}=(\exp(Z^{*}_1), \dots,\exp(Z^{*}_n))'$ and ${\ln {\mathbf{Z}}^*}=(\ln Z^{*}_1, \dots,\ln Z^{*}_n)'$.

\begin{Definition} (Log-CFUSN family of  distributions)
\label{Def1}
Let $\bZ^{*}$ and $\bY$ be $n \times 1$ random vectors such that $\bZ^{*} = \ln \bY$. We say
that $\bY $  has a log-canonical-fundamental-skew-normal distribution with $ n \times m$ skewness matrix 
$\mathbf{\Delta}$ denoted by $\mathbf{Y} \sim LCFUSN_{n,m}(\mathbf{\Delta})$, if
$\mathbf{Z^{*}} \sim CFUSN_{n,m}(\mathbf{\Delta})$ with pdf given in (\ref{CFUSN}).
\end{Definition}

Thus, from definition \ref{Def1}, we have that $\mathbf{Y}={\exp({\mathbf{Z}}^*)}$
and  using some results of probability calculus, we can prove that the pdf of
the log-CFUSN family of  distributions with skewness matrix 
$\mathbf{\Delta}$ is
\begin{equation}\label{LCFUSN}
f_{\mathbf{Y}}(\mathbf{y})=2^{m}\left(\prod_{i=1}^{n}y_{i}\right)^{-1}
\phi_{n}(\ln \mathbf{y})\Phi_{m}(\mathbf{\Delta}'\ln \mathbf{y}|{\bf{I}}_{m}-\mathbf{\Delta}'\mathbf{\Delta}),
\,\,\,\,\mathbf{y} \in \mathbb{R}^{n^{+}},
\end{equation}
where $\mathbf{\Delta}$ is an $n \times m$ matrix such that $||\mathbf{\Delta} \textbf{a}||<1$, for all unity vectors
\textbf{a} $\in \mathbb{R}^{m}$. 


This distribution generalizes the multivariate log-SN distribution  defined by  \cite{MaGe10} by assuming a $m$-variate skewing function.
If in (\ref{LCFUSN}) we take $m=1$ and assume $\mathbf{\alpha}=
({\bf{I}}_{m}-\mathbf{\Delta}'\mathbf{\Delta})^{-\frac{1}{2}}\mathbf{\Delta}'$ we obtain the family defined by  \cite{MaGe10}
which general  expression is given in $(\ref{logskewelliptical})$. If $\mathbf{\Delta}$ is a
matrix with all entries equal to zero we have the multivariate log-normal distribution. Another reason to study this distribution comes from
results in \cite{SaLoAr13} summarized in the introduction. As it can be noticed, the distribution for the
odds ratio given in (\ref{EqLNA11}) also belongs to the log-CFUSN family of  distributions whenever
the individuals under comparison have the same characteristics, that is, equal vector of covariates
($\bx^{t}_{i_1j_1}=\bx^{t}_{i_2j_2}$), and the scale parameter for the distribution of the random effects
is $\sigma^2 =1$.  In that case, $OR \sim LCFUSN_{1,2}(\mathbf{\Delta})$ where $ \mathbf{\Delta}= \delta \bepsilon$.

\begin{figure}[h!]
  \centering
   \includegraphics[width=7cm]{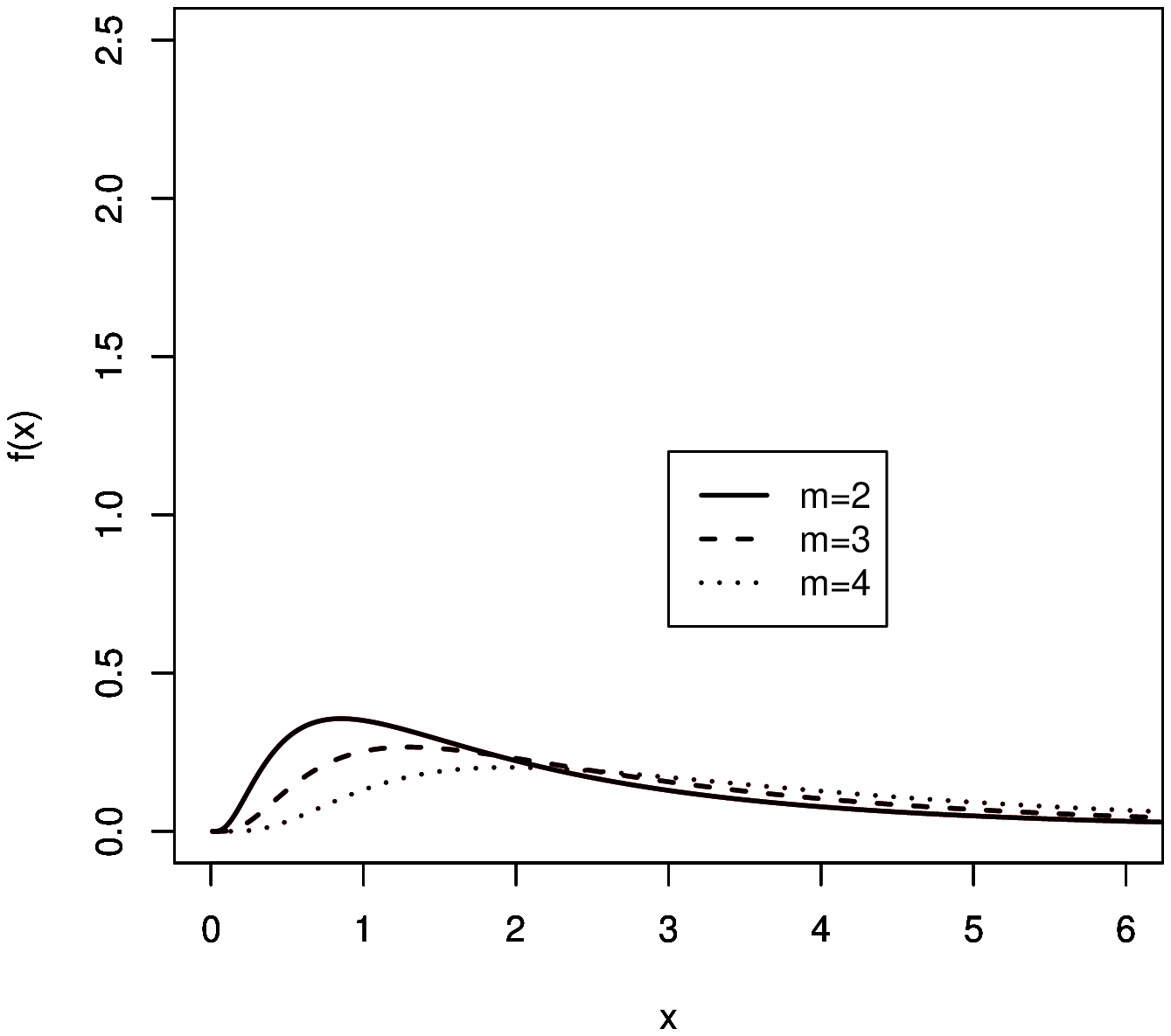}  
   \includegraphics[width=7cm]{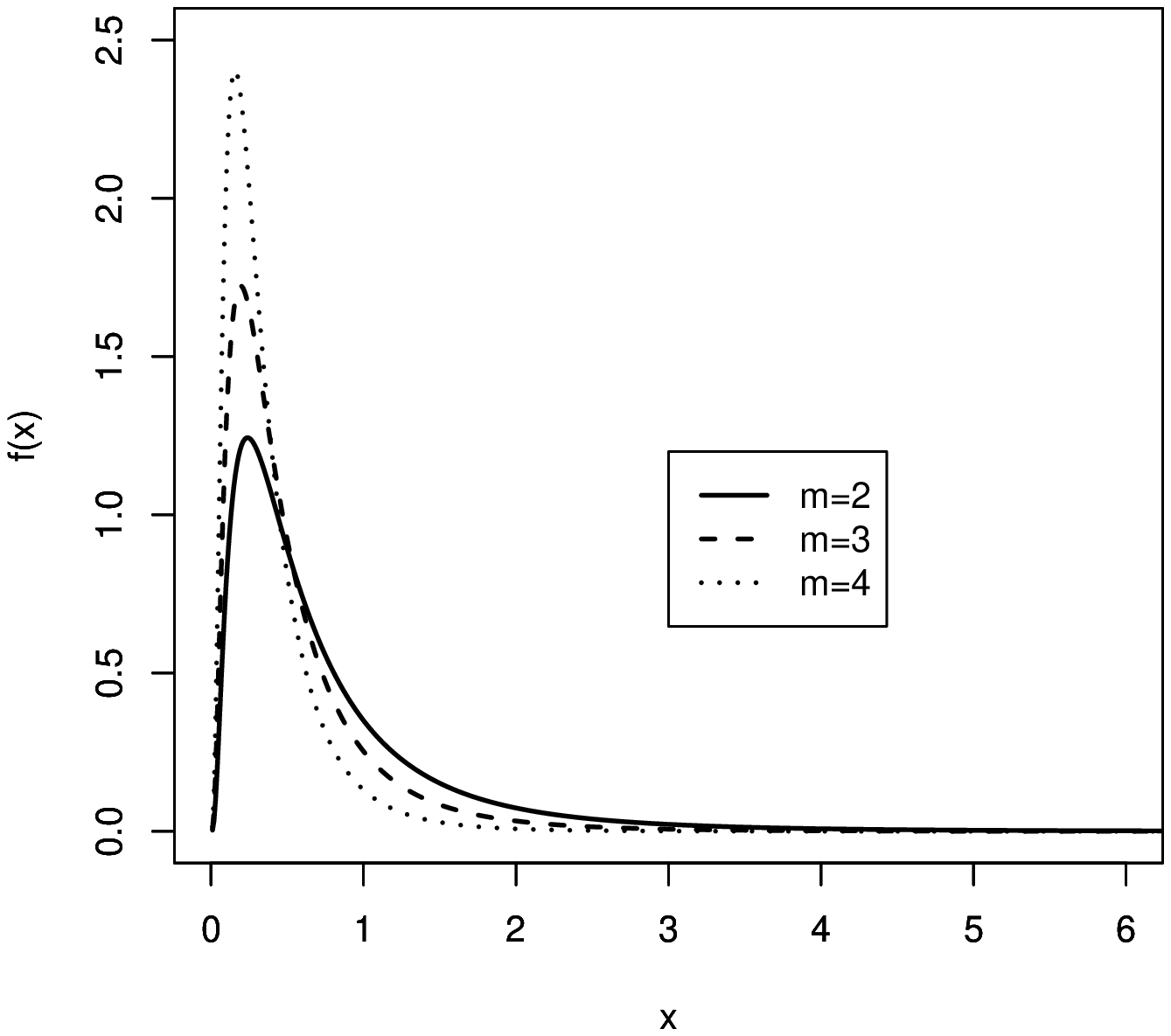}
  \caption{Log-CFUSN densities $LCFUSN_{1,m}(\mathbf{\Delta})$ for different
  values of $m$ and $\mathbf{\Delta}=0.4 \times{\bf{1}}_{m}'$ (left) and $\mathbf{\Delta}= -0.4\times{\bf{1}}_{m}'$ (right).}
  \label{densidade}
\end{figure}

Figure \ref{densidade} depicts the densities of $LCFUSN_{n,m}(\mathbf{\Delta})$ for the case $n=1$ and some values
of $m$ and $\mathbf{\Delta}$. To simplify  the presentation let  ${\bf{1}}_{n,m}$ be the matrix of ones of order
$n \times m$ and denote by ${\bf{1}}_{n}$ the column vector of ones of order $n$. Clearly the distribution allocates more mass to the tails when $m$ increases. Moreover,
the densities shape becomes more flexible if compared with  (\ref{logskewelliptical}).  

\begin{figure}[htb!]
  \centering
  \includegraphics[width=13cm]{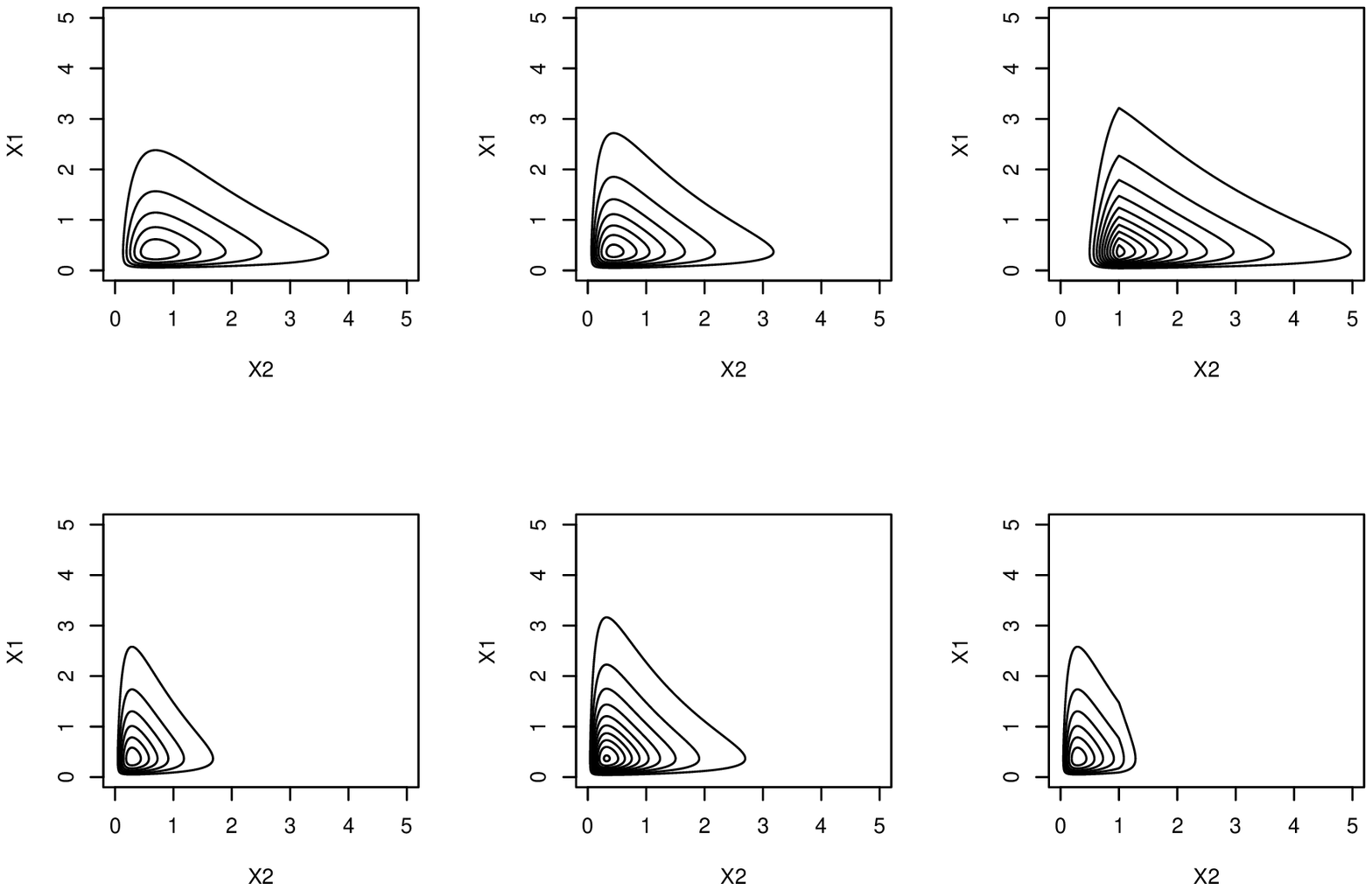}\\
  \caption{Contour plots for the log-CFUSN densities with $n=m=2$ and $\mathbf{\Delta_1}$ (top left), $\mathbf{\Delta_2}$ (top middle), $\mathbf{\Delta_3}$ (top right), $\mathbf{\Delta_4}$ (bottom left), $\mathbf{\Delta_5}$ (bottom middle), $\mathbf{\Delta_6}$ (bottom right).}
  \label{cnivelm2}
  \centering
  \includegraphics[width=13cm]{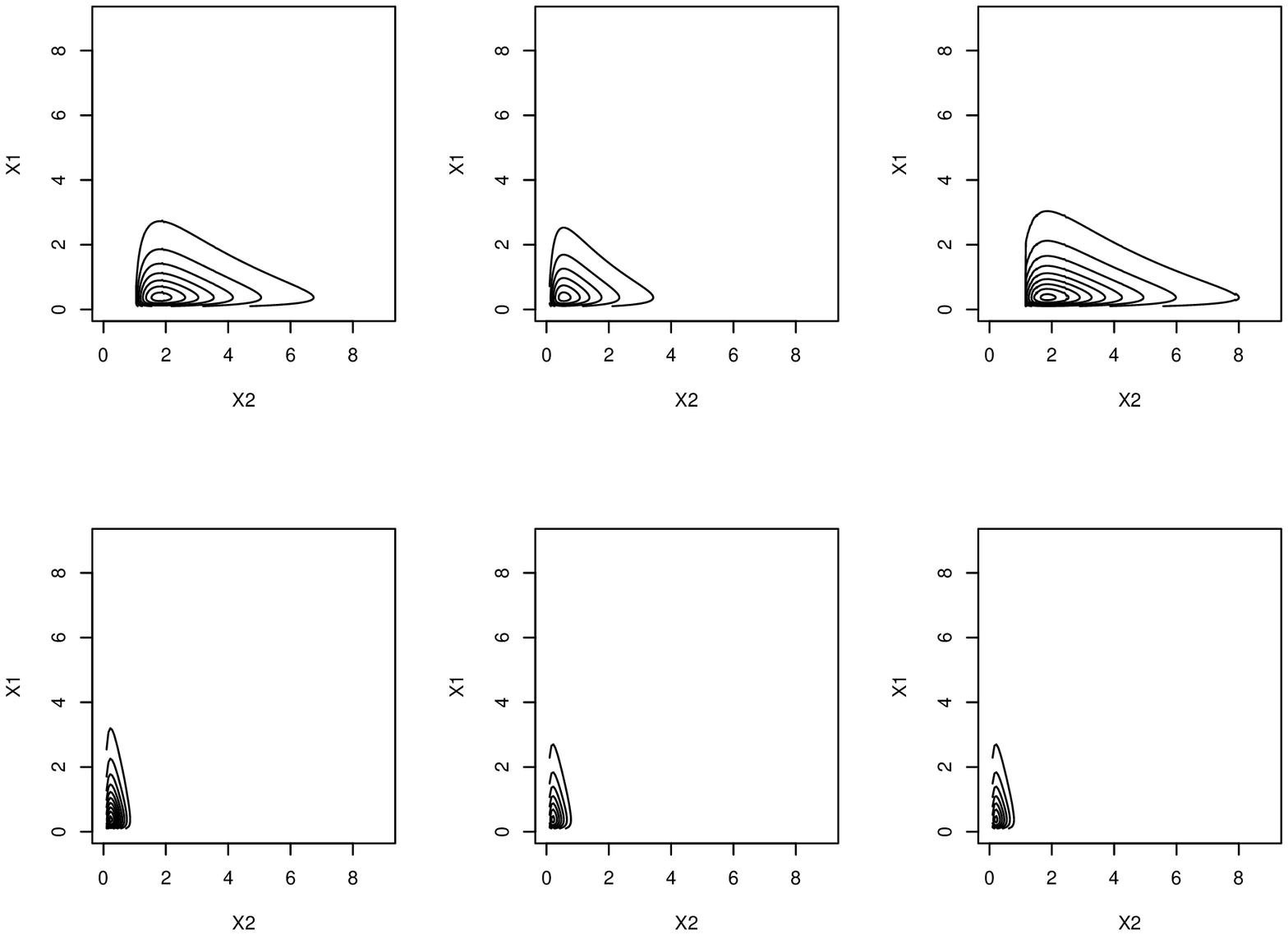}\\
  \caption{Contour plots for the log-CFUSN densities with $n=2$ and $m=3$ and $\mathbf{\Delta_1}$ (top left), $\mathbf{\Delta_2}$ (top middle), $\mathbf{\Delta_3}$ (top right), $\mathbf{\Delta_4}$ (bottom left), $\mathbf{\Delta_5}$ (bottom middle), $\mathbf{\Delta_6}$ (bottom right).}
  \label{cnivelm3}
\end{figure}

In order to show the effect of $m$ in the asymmetry of the distribution, Figures \ref{cnivelm2}
and \ref{cnivelm3}  show the contour plots for the log-CFUSN densities $LCFUSN_{n,m}(\mathbf{\Delta})$
whenever $m=2$ and $3$, respectively. In both cases we assume bivariate ($n=2$) log-CFUSN densities.
In Figure \ref{cnivelm2} the following skewness matrices of parameters $\mathbf{\Delta}$ are assumed
$\mathbf{\Delta}_1= - \mathbf{\Delta}_4 = 0.3 \times {\bf{1}}_{2,2}$,
$\mathbf{\Delta}_2= - \mathbf{\Delta}_5= 0.1 \times {\bf{1}}_{2,2}$ and 
$\mathbf{\Delta}_3= - \mathbf{\Delta}_6 =\left(\begin{array}{cc}0.4 & 0.8 \\0.3 & 0.3\end{array}\right) $.
%
%
In Figure \ref{cnivelm3} the skewness matrices of parameters $\mathbf{\Delta}$ are
$\mathbf{\Delta}_1= - \mathbf{\Delta}_4 = 0.3 \times {\bf{1}}_{2,3}$,
$\mathbf{\Delta}_2= - \mathbf{\Delta}_5= 0.2 \times {\bf{1}}_{2,3}$, $\mathbf{\Delta}_3=0.1 \times {\bf{1}}_{2,3}$ and 
%
 $\mathbf{\Delta}_6=\left(\begin{array}{ccc}-0.1 & -0.3 & -0.2 \\-0.1 & -0.3 & -0.2\end{array}\right)$.

It is clear that the curves in Figures \ref{cnivelm2} and \ref{cnivelm3} deviate from the origin when the entries
of $\mathbf{\Delta}$ are positive and curves are more concentrated around the origin when these
entries are negative. Similar behavior is noted in the contour curves of the $CFUSN_{n,m}(\bDelta)$ distribution
in \cite{ArGe05}.
\clearpage

\clearpage

It must be also noticed that the log-CFUSN family of distributions is a subclass of an extended class of log-skewed
distributions with normal kernel which  can be built similarly from the family 
defined by \cite{ArAz06}. If we consider the SUN family of distribution in (\ref{DeSUN}), we can define the log-SUN
family of distibution as follows. 

\begin{Definition} (Log-SUN family of  distributions)
\label{DefLSUN}
Let $\bZ^{*}$ and $\bY$ be $n \times 1$ random vectors such that $\bZ^{*} = \ln \bY$. We say
that $\bY $  has a log-unified-skew-normal distribution with parameters
$\bfeta$, $\bgamma$, $\bar{\bomega}$ and  $\bOmega^*$ as defined in
(\ref{DeSUN}) denoted by $\mathbf{Y} \sim LSUN_{n,m}(\bfeta,\bgamma, \bar{\bomega}, \bOmega^*)$, if
$Z^* \sim SUN_{n,m}(\bfeta,\bgamma, \bar{\bomega}, \bOmega^*)$  with pdf given in (\ref{DeSUN}).
\end{Definition}
It follows, as a consequence of Definition \ref{DefLSUN}, that the pdf of $\bY $ is given by
\begin{equation}
\label{LogSUN}
f_{\bY}(\mathbf{y})= \left(\prod_{i=1}^{n}y_{i}\right)^{-1}
\phi_n(\ln{\by}- \bfeta\mid \bOmega)
\frac{\Phi_m( \bgamma + \bDelta'\bar{\bOmega}^{-1}\bomega^{-1}(\ln{\by}- \bfeta)|\bGamma - \bDelta'\bar{\bOmega}^{-1} \bDelta)}
{\Phi_m(\bgamma \mid \bGamma)},\,\,\,\, 
\end{equation}
for $\mathbf{y}\in\RR^n_{+}.$

Particularly,  if $\mathbf{Y} \sim LSUN_{n,m}(\bf{0},\bf{0}, {\bf{1}}_n, \bOmega^*)$, where ${\bf{1}}_n$ is the
column vector of ones of order $n$  and
$
\bOmega^* = \left(
\begin{array}{cc}
\bI_m & \bDelta' \\
\bDelta & \bI_n \\
\end{array}\right),
$
it follows that $\mathbf{Y} \sim LCFUSN_{n,m}(\mathbf{\Delta})$ with pdf given in (\ref{LCFUSN}).

\subsection{Some properties of the Log-CFUSN family of distributions}

We now present several properties of the log-CFUSN family of distributions, among them are the
mixed moments, the cdf and, marginal and conditional distributions.
We also establish conditions for independence in the log-CFUSN family of distributions.
Proposition \ref{propcfusncdf} provides the cdf for this family.

\begin{prop}
\label{propcfusncdf}
If $\mathbf{Y} \sim LCFUSN_{n,m}(\mathbf{\Delta})$, then its cdf is given by
\begin{equation}\label{fdalcfusn}
F_{\mathbf{Y}}(\mathbf{y})=2^{m}\Phi_{n+m}((\ln \mathbf{y'},\mathbf{0}')'|\mathbf{\Omega}), \;\;\mathbf{y} \in \mathbb{R}^{n^{+}}
\end{equation}
where $\mathbf{\Omega}=\left(
\begin{array}{cc}
\mathbf{I}_{n} & -\mathbf{\Delta} \\
-\mathbf{\Delta}' & \mathbf{I}_{m} \\
 \end{array}\right).$
\end{prop}

The proof of Proposition \ref{propcfusncdf} follows from Proposition 2.1 in \cite{ArGe05}
by noticing that $P(\mathbf{Y}\leq \mathbf{y})=P(\exp({\mathbf{Z}^{*})} \leq \mathbf{y})=P(\mathbf{Z}^{*} <\ln \mathbf{y})=
F_{\mathbf{Z}^{*}}(\ln \mathbf{y})$.

The mixed moments of a random vector $\mathbf{Y}\sim LCFUSN_{n,m}(\mathbf{\Delta})$ can be expressed in terms of
the moment generating function of a $CFUSN_{n,m}(\mathbf{\Delta})$ distribution. This can be seen in the following proposition.

\begin{prop}
\label{momentoslcfusn}
If $ \mathbf{Y} \sim LCFUSN_{n,m}(\mathbf{\Delta})$ and $\mathbf{t}=(t_{1}, ..., t_{d})'$, $t_{i}
\in \mathbb{N}$, then the mixed moments of $ \mathbf{Y}$ are given by
\begin{equation}\label{mmlcfusn}
E(\prod_{i=1}^{n}{Y_{i}}^{t_{i}})=2^{m}e^{(1/2)\mathbf{t}'\mathbf{t}}\Phi_{m}(\mathbf{\Delta}'\mathbf{t}).
\end{equation}
\end{prop}

The proof of Proposition \ref{momentoslcfusn} follows by noticing that $E(\prod_{i=1}^{n}{Y_{i}}^{t_{i}})$ $=$
$E(\prod_{i=1}^{n}e^{{t_{i}}\ln{Y_{i}}})$ $=$
$E(e^{\sum_{i=1}^{n}{t_{i}}\ln{Y_{i}}})$ $= $ $E(e^{\mathbf{t} \ln \mathbf{Y}})$ $=$ $M_{\ln \mathbf{Y}}(\mathbf{t})$.
As $ \mathbf{Y} \sim LCFUSN_{n,m}(\mathbf{\Delta})$ , we have
$ \ln \mathbf{Y} \sim CFUSN_{n,m}(\mathbf{\Delta})$. The result follows from Proposition 2.3 in \cite{ArGe05}.

Considering the result in $(\ref{mmlcfusn})$, we can calculate the moments of a random vector with distribution
$LCFUSN_{n,m}(\mathbf{\Delta})$.  For example, if we consider $\mathbf{Y} \sim LCFUSN_{1,m}(\mathbf{\Delta})$, we
have that 
\begin{eqnarray}
  E(Y) &=& 2^{m}e^{1/2}\Phi_{m}(\bDelta)\nonumber\\ \nonumber
  E(Y^{2}) &=& 2^{m}e^{2}\Phi_{m}(2\bDelta) \\ \nonumber
  E(Y^{3}) &=& 2^{m}e^{9/2}\Phi_{m}(3\bDelta) \\ \nonumber
  E(Y^{4}) &=& 2^{m}e^{8}\Phi_{m}(4\bDelta) \nonumber.
\end{eqnarray}

Considering these results it can be proved that   the coefficient of  asymmetry and kurtosis of $Y\sim  LCFUSN_{1,m}(\bDelta)$  are given, respectively, by
%
\begin{equation}\label{asymmetrylcfusn}
\gamma_{Y}= \frac{e^{3}\Phi_{m}(3\bDelta)-2^{m}\Phi_{m}(\bDelta)(3e\Phi_{m}(2\bDelta)-2\Phi_{m}(\bDelta))}{2^{\frac{m}{2}}(e \Phi_{m}(2\bDelta)-2^{m}\Phi^{2}_{m}(\bDelta))^{3/2}},
\end{equation} and
\begin{equation}\label{kurtosislcfusn}
\kappa_{Y}= \frac{e^{6}\Phi_{m}(4\bDelta)-2^{m}(4e^{3}\Phi_{m}(\bDelta)\Phi_{m}(3\bDelta)-3.2^{m+1}e\Phi^{2}_{m}(\bDelta)\Phi_{m}(2\bDelta)+3.2^{2m}\Phi^{4}_{m}(\bDelta))}{2^{m}(e^{2}\Phi^{2}_{m}(2\bDelta)-2^{m+1}e\Phi_{m}(2\bDelta)\Phi^{2}_{m}(\bDelta)+2^{2m}\Phi^{4}_{m}(\bDelta))}.
\end{equation}
Consequently, if   $m=1$  and $\Delta$  is a matrix with all entries equal to zero, that is, if  $Y \sim LN(0,1)$ then
$\gamma_{Y}= (2+e)\sqrt{e-1}$  and  $\kappa_{Y}=e^{4}+2e^{3}+3e^{2}-3$.

\begin{figure}[h]                                                                                                                                                   
\centering                                                                                                                                                         
\includegraphics[width=12cm]{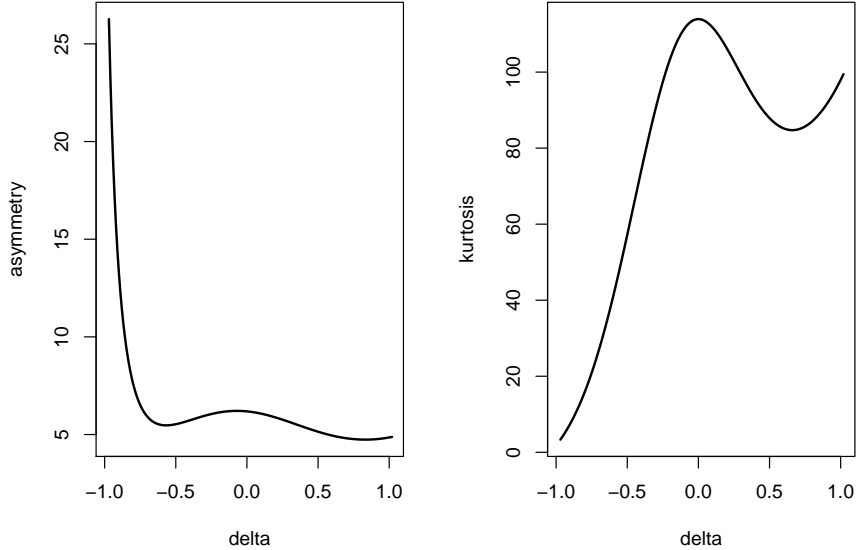}\\
\caption{Asymmetry (left) and Kurtosis (right) for the $LCFUSN_{1,1}(\bDelta)$ distribution.} \label{asykurt}
\end{figure}

Figure \ref{asykurt} depicts the asymmetry coefficient and kurtosis for the $LCFUSN_{1,1}(\bDelta)$ distribution. Observe that $\bDelta=0$ corresponds to the log normal case. It is clear, at least in the case $n=m=1$, that asymmetry and kurtosis can change significantly depending on the choice of $\bDelta$. 

Table \ref{kurtsym} displays the asymmetry and kurtosis coefficients of the $LCFUSN_{1,m}(\bDelta)$ 
as a function of $m$  and it suggests a monotonic decreasing behavior  of these quantities as $m$ increases. 
Although the behavior of these coefficients depends on $\bDelta$, particularly,
for $\bDelta=0.4\times{\bf{1}}_{m}'$ and  $\bDelta=-0.4\times{\bf{1}}_{m}'$ 
the asymmetry and kurtosis coefficients  of the $LCFUSN_{1,m}(\Delta)$ 
are both smaller than those obtained for the $LN(0,1)$  for all $m$ considered in the study.

\clearpage

\begin{table}[htb]
\begin{center}
\footnotesize
\caption{Kurtosis and asymmetry for the $LCFUSN_{1,m}(\bDelta)$. }
\label{kurtsym}
\tabcolsep=3pt 
\begin{tabular}{cccccccc}
\hline
\multicolumn{1}{c}{}&& \multicolumn{3}{c}{$\bDelta=0.4\times{\bf{1}}_{m}'$} &\multicolumn{3}{c}{$\bDelta=-0.4\times{\bf{1}}_{m}'$} \\
\cline{3-4}\cline{6-7}  
$m$ & &Kurtosis & Asymmetry & & Kurtosis & Asymmetry. &  \\ \hline
1 &&  $92.84$ & $5.64$ &&  $74.39$ &  $5.20$ &  \\
2 &&  $76.30$ & $5.16$ &&  $48.38$ &  $4.33$ & \\
3 &&  $63.39$ & $4.73$ &&  $31.12$ &  $3.55$ &  \\
4 &&  $53.42$ & $4.36$ &&  $19.52$ &  $2.84$ & \\
5 &&  $45.91$ & $4.05$ &&  $11.59$ &  $2.14$ & \\
\hline
\end{tabular}
\end{center}
\end{table}

Similar to what is observed for the CFUSN family of distributions, the log-CFUSN is closed  under marginalization  but  not
under conditioning. The next result establishes that the $LCFUSN_{n,m}(\mathbf{\Delta})$ distribution is closed under marginalization.
The proof of this result will be omitted. It follows immediately from Proposition 2.6 in \cite{ArGe05} and Definition \ref{Def1}.
  
\begin{prop}
\label{marginallcfusn}
Let $\mathbf{Y}\sim LCFUSN_{n,m}(\mathbf{\Delta})$ and consider the partitions $\mathbf{Y}=\left(
\mathbf{Y}_{1}', \mathbf{Y}_{2}'  \right)'$ and $\mathbf{\Delta}=\left(\mathbf{\Delta}_{1}',
\mathbf{\Delta}_{2}' \right)'$, where  $\mathbf{Y}_{i}$  and $\mathbf{\Delta}_{i}$
has dimensions $n_{i} \times 1$ and $n_{i}\times m$, respectively,  and $n_{1}+n_{2}=n$. Then, for $i=1,2$, $\mathbf{Y}_{i}\sim LCFUSN_{n_{i},m}(\mathbf{\Delta}_{i})$ with pdf given by

\begin{equation}
\label{fdplcfusnmarg}
f_{\mathbf{Y}_{i}}(\mathbf{y}_{i})=2^{m}\left(\prod_{j=1}^{n_{i}}y_{j}\right)^{-1}\phi_{n_{i}}(\ln \mathbf{y}_{i})\Phi_{m}(\mathbf{\Delta}'_{i}\ln \mathbf{y}_{i}|\mathbf{I}_{m}-\mathbf{\Delta}'_{i}\mathbf{\Delta}_{i}), \mathbf{y}_{i} \in \mathbb{R}^{n_{i}^{+}}.
\end{equation}
\end{prop}

It is also possible to derive conditions for independence under the log-CFUSN family of distributions by assuming some
constraints on the partitions defined in Proposition \ref{marginallcfusn}.

\begin{prop}
\label{indlcfsun}
Let $\mathbf{Y} \sim LCFUSN_{n,m}(\mathbf{\Delta})$ and consider the partitions $\mathbf{Y}=\left(
\mathbf{Y}_{1}', \mathbf{Y}_{2}'  \right)'$ and $\mathbf{\Delta}=\left(\mathbf{\Delta}_{1}',
\mathbf{\Delta}_{2}' \right)'$, where  $\mathbf{Y}_{i}$  and $\mathbf{\Delta}_{i}$
has dimensions $n_{i} \times 1$ and $n_{i}\times m$, respectively, and $n_{1}+n_{2}=n$. Let $\bDelta_i=(\bDelta_{i,1},\bDelta_{i,2})$, where
$\bDelta_{i,j}$ has dimension $n_i \times m_j$, $j=1,2$, and $m_1+m_2=m$, $m>1$.
Then, under each  of the conditions below on the shape matrix $\mathbf{\Delta}$, the random vectors
$\mathbf{Y}_{1}$ and $\mathbf{Y}_{2}$ are independent
\begin{itemize}
\item[(i)] $\mathbf{\Delta}_{12}=\mathbf{\Delta}_{21}=\mathbf{0}$ and, in this case, $\mathbf{Y}_{i}\sim LCFUSN_{n_{i},m_{i}}(\mathbf{\Delta}_{ii}), i=1,2$;\\
\item[(ii)] $\mathbf{\Delta}_{ii}=\mathbf{0}, i=1,2$ and, in this case, $\mathbf{Y}_{1}\sim LCFUSN_{n_{1},m_{2}}(\mathbf{\Delta}_{12})$ e $\mathbf{Y}_{2}\sim LCFUSN_{n_{2},m_{1}}(\mathbf{\Delta}_{21})$.
\end{itemize}
\end{prop}

The proof of Proposition \ref{indlcfsun} is straightforward from Proposition 2.7 in \cite{ArGe05}
and thus is omitted. We now obtain the conditional distributions under the $LCFUSN_{n,m}(\bDelta)$ family.

\begin{prop}
\label{lcfusncond}
Let $\mathbf{Y}\sim LCFUSN_{n,m}(\mathbf{\Delta})$ and consider the partitions $\mathbf{Y}=\left(
\mathbf{Y}_{1}', \mathbf{Y}_{2}'  \right)'$ and $\mathbf{\Delta}=\left(\mathbf{\Delta}_{1}',
\mathbf{\Delta}_{2}' \right)'$, where
$\mathbf{Y}_{i}$ and $\mathbf{\Delta}_{i}$ has dimensions $n_{i} \times 1$ and $n_{i}\times m$, respectively,
and $n_{1}+n_{2}=n$. Then, the conditional pdf of $\mathbf{Y}_{1}$ given $\mathbf{Y}_{2}=\mathbf{y}_{2}$,
$\mathbf{y}_{2}\in \mathbb{R}^{n_{2}^{+}}$ is given by
\begin{equation}
\label{lcfusncondeq}
f_{\mathbf{Y}_{1}|\mathbf{Y}_{2}=\mathbf{y}_{2}}(\mathbf{y}_{1})=\left(\prod_{i=j}^{n_{1}}y_{j}\right)^{-1}\phi_{n_{1}}(\ln \mathbf{y}_{1})
\frac{\Phi_{m}(\mathbf{\Delta}_{1}'\ln \mathbf{y}_{1}|-\mathbf{\Delta}'_{2}\ln \mathbf{y}_{2},\mathbf{I}_{m}-\mathbf{\Delta}'\mathbf{\Delta})}{\Phi_{m}(\mathbf{\Delta}'_{2}\ln \mathbf{y}_{2}|\mathbf{I}_{m}-\mathbf{\Delta}'_{2}\mathbf{\Delta}_{2})}, \;\;\mathbf{y}_{1} \in \mathbb{R}^{n_{1}^{+}}.
\end{equation}
\end{prop}

The proof follows from results of probability calculus and by
noticing that, given $\mathbf{y}_{2} \in \mathbb{R}^{n_{2}^{+}}$, we have that
$\Phi_{m}(\mathbf{\Delta}'\ln{\by}|\mathbf{I}_{m} - \mathbf{\Delta}'\mathbf{\Delta})=$
$\Phi_{m}(\mathbf{\Delta}_{1}'\ln{\by}_{1}+\mathbf{\Delta}_{2}'\ln{\by}_{2}|\mathbf{I}_{m} -
\mathbf{\Delta}'\mathbf{\Delta})=\Phi_{m}(\mathbf{\Delta}_{1}'\ln{\by}_{1}|-\mathbf{\Delta}_{2}'\ln{\by}_{2},
\mathbf{I}_{m} - \mathbf{\Delta}'\mathbf{\Delta})$.

Notice  that the log-CFUSN family of distribution per se is not closed under conditioning. However,
if considered as a particular subclass of the log-SUN family of distribution, we notice
from (\ref{lcfusncondeq}) and (\ref{LogSUN}) that
${\mathbf{Y}}_{1} \mid {\mathbf{Y}}_{2}={\mathbf{y}}_{2} \sim
LSUN_{n,m}({\bf{0}},\bDelta_{2}'\ln {\mathbf{y}}_{2}, {\bf{1}}_{n_1}, \bOmega^*)$, where
$
\bOmega^* = \left(
\begin{array}{cc}
\bI_m - \bDelta_2'\bDelta_2 & \bDelta_1' \\
\bDelta_1 & \bI_{n_1} \\
 \end{array}\right).
$

\subsection{A location-scale extension of the log-CFUSN distribution}

More flexible class of distributions are obtained if we are able to
include on it  location and scale parameters. Usually, this is done considering a linear
transformation of a variable with the standard distribution. Assuming
this principle, we introduce the location-scale extension of the $LCFUSN_{n,m}$ distribution
as follows. 

Assume that $\mathbf{X}\sim CFUSN_{n,m}(\mathbf{\Delta})$ and define the linear transformation $\mathbf{W}={\bmu}+
\mathbf{\Sigma}^{\frac{1}{2}}\mathbf{X}$, where $\bmu$  is an $n \times 1$ vector  and  $\mathbf{\Sigma}$ is
an $n \times n$ positive definite matrix. As shown by \cite{ArGe05}, the pdf of $\mathbf{X}$ is 
\begin{equation}\label{fdp2}
f_{\mathbf{W}}(\mathbf{w})=2^{m}|\mathbf{\Sigma}|^{-1/2}\phi_{n}(\mathbf{\Sigma}^{-1/2}(\mathbf{w}-{\bmu}))
\Phi_{m}(\mathbf{\Delta}'\mathbf{\Sigma}^{-1/2}(\mathbf{w}-{\bmu})|\mathbf{I}_{m}-\mathbf{\Delta}'\mathbf{\Delta}),
\mathbf{w} \in \mathbb{R}^{n}. 
\end{equation}
Let us  consider the transformation $\mathbf{U}= \exp({\mathbf{W}})$. By definition,
$\mathbf{U}$  has a location-scale  log-CFUSN distribution denoted by $\mathbf{U}
\sim LCFUSN_{n,m}({\bmu}, \mathbf{\Sigma}, \mathbf{\Delta})$ and its  pdf is
\begin{eqnarray}
\label{fdpescalaloc}
f_{\mathbf{U}}(\mathbf{u})&=&2^{m}|\mathbf{\Sigma}|^{-1/2}\left(\prod_{j=1}^{n}u_{j}\right)^{-1}
\phi_{n}(\mathbf{\Sigma}^{-1/2}(\ln \mathbf{u}-{\bmu})) \nonumber \\
& \times &\Phi_{m}(\mathbf{\Delta}'\mathbf{\Sigma}^{-1/2}
(\ln \mathbf{u}-{\bmu})|\mathbf{I}_{m}-\mathbf{\Delta}'\mathbf{\Delta}), \mathbf{u} \in \mathbb{R}^{n^{+}}.
\end{eqnarray}

It is important to note that if $\Sigma = {\rm diag}\{\sigma_1^2, \dots, \sigma_n^2\}$,
that is, if we are skewing an independent $n$-variate  normal distribution,
the distribution in (\ref{fdpescalaloc}) can be obtained from the
log-SUN distribution given in (\ref{LogSUN}) by assuming $\bfeta = \mu$, $\bgamma = \bf{0}$, $\bar{\bomega} =
(\sigma_1, \dots, \sigma_n)$ and
$
\bOmega^* = \left(
\begin{array}{cc}
\bI_m & \bDelta' \\
\bDelta & \bI_n \\
 \end{array}\right),
$
that is, we have that $\mathbf{\mathbf{U}}  \sim LSUN_{n,m}(\bmu,\bf{0}, \bar{\bomega}, \bOmega^*)$.

Marginal and conditional distributions in the location-scale log-CFUSN class of distributions
are not easily obtainable. However, under some particular
structures for $\mathbf{\Sigma}$ we can derive such results.
Let $W \sim CFUSN_{n,m}(\mathbf{\mu}, \mathbf{\Sigma}, \mathbf{\Delta})$,  as defined in Expression 2.11
in \cite{ArGe05}, and  consider the partitions $$\mathbf{W}=\left(
\begin{array}{c}
\mathbf{W}_{1} \\
\mathbf{W}_{2} \\
\end{array}
\right), \mathbf{\Delta}=\left(
\begin{array}{c}
\mathbf{\Delta}_{1} \\
\mathbf{\Delta}_{2} \\
\end{array}
\right), \bmu=\left(
 \begin{array}{c}
{\bmu}_{1} \\
{\bmu}_{2} \\
\end{array}
\right),$$ where $\mathbf{W}_{i}$, ${\bmu}_{i}$ and $\mathbf{\Delta}_{i}$ have dimensions $n_{i}\times 1$, $n_{i}\times 1$ and
$n_{i}\times m$, $i=1,2$, respectively, and $n_{1}+n_{2}=n$. Suppose also that $\mathbf{\Sigma}$  is a diagonal matrix such that
 $$\mathbf{\Sigma} = \left(
\begin{array}{cc}
\mathbf{\Sigma}_{11} & \textbf{0}  \\
\textbf{0}  & \mathbf{\Sigma}_{22}  \\
\end{array}
\right),$$ where $\mathbf{\Sigma}_{ij}$ has dimension $n_{i}\times n_{j}$.  Under these conditions,
it follows that   $\mathbf{U}_i= \exp({\mathbf{W}_{i}})
\sim LCFUSN({\bmu}_i, \mathbf{\Sigma}_{ii}, \mathbf{\Delta}_i)$, that is the location-scale
log-CFUSN family of distributions preserves closeness under marginalization.
                                                                                                                                                                                                                                         
It also follows that the conditional distribution of $\mathbf{U}_{1}|\mathbf{U}_{2}=\mathbf{u}_{2}$ is given by                                                                                                                                                                                       
\begin{eqnarray}
\label{condcomescala1}                                                                                                                                                                                                                      
f_{\mathbf{U}_{1}|\mathbf{U}_{2}=\mathbf{u}_{2}}(\mathbf{u}_{1})&=& \left(\prod_{j=1}^{n_{1}}u_{j}\right)^{-1}
\phi_{n_{1}}(\mathbf{\Sigma}_{11}^{-1/2}(\ln \mathbf{u}_{1}-{\bmu}_{1})) \\
&\times& \frac{\Phi_{m}(\mathbf{\Delta}_{1}'\mathbf{\Sigma}_{11}^{-1/2}
(\ln \mathbf{u}_{1}-{\bmu}_{1})|-\mathbf{\Delta}_{2}'\mathbf{\Sigma}_{22}^{-1/2}(\ln \mathbf{u}_{2}-{\bmu}_{2}), I_{m}-\mathbf{\Delta}'\mathbf{\Delta})}
{\Phi_{m}(\mathbf{\Delta}_{2}'\mathbf{\Sigma}_{22}^{-1/2}(\ln \mathbf{u}_{2}-{\bmu}_{2})|I_{m}-
\mathbf{\Delta}_{2}'\mathbf{\Delta}_{2})},  \nonumber                                                                                                                                                                                                                               
\end{eqnarray}                                                                                                                                                                                                                                             
$\mathbf{u}_{1} \in \mathbb{R}^{n_{1}^{+}}$ and $\mathbf{u}_{2} \in \mathbb{R}^{n_{2}^{+}}$.

\subsection{Stochastic representation}

Stochastic representations of skewed distributions are useful, for instance, to  generate samples
from those distributions more easily. They also  play a very  important role in inference
if we are interested in apply MCMC or EM methods.

A stochastic representation of the log-CFUSN family is straightforward  from the 
marginal  stochastic representation of the CFUSN family given
in  \cite{ArGe05}.

Assume that $\bZ^{*}\sim CFUSN_{n,m}(\bDelta)$, where $||\bDelta'\ba||<1$ for any unitary vector $\ba \in \RR^n$.
Let $\bD \sim N_{m}(\bf{0}, \bI_{m})$, $\mathbf{V} \sim N_{n}(\bf{0},\bI_{n})$ where $\bD$ and $\mathbf{V}$ are
independent column random vectors of order $m$ and $n$, respectively. Denote by $|\bD|$ the vector $(|D_{1}|,...,|D_{m}|)'$.
\cite{ArGe05} prove that the marginal representation of $\bZ^{*}$ is
\begin{equation}
\label{EqSR}
\bZ^{*}\buildrel d \over = \bDelta |\bU|+(\bI_{n}-\bDelta\bDelta')^{1/2}\mathbf{V}.
\end{equation}

If $\bY \sim LCFUSN_{n,m}(\bDelta)$ then its  marginal representation  follows  as a consequence of  (\ref{EqSR}) by noticing that  
$\bY \buildrel d \over = \exp({\bZ^*})$ $\buildrel d \over = \exp(\bDelta |\bD|) \exp((\bI_{n}-\bDelta\bDelta')^{1/2}\mathbf{V})$
$\buildrel d \over = \exp(\bDelta |\bD|) \bT$, where $\bT$ has a multivariate log-normal distribution
with a null location parameter and scale matrix equal to $\bI_{n}-\bDelta\bDelta'$.

\section{Some aspects of Bayesian Inference in the LCFUSN Family}
\label{Sec3}

Let $\bY_1, \dots, \bY_L \mid \bmu, \bSigma, \bDelta \buildrel iid \over\sim LCFUSN_{n,m}(\bmu, \bSigma, \bDelta)$  with
pdf given in  (\ref{fdpescalaloc}). Define the $L \times n$ matrices $\bY= (\bY_1, \dots, \bY_L)'$
and  $\ln\bY= (\ln\bY_1, \dots, \ln\bY_L)'$.  Therefore, it follows that the likelihood function
is given by
\begin{eqnarray}
\label{EqVeroG1}
f(\bY \mid  \bmu, \bSigma, \bDelta) &=& 2^{Lm} \prod_{l=1}^{L} \prod_{j=1}^{n} Y_{lj}^{-1}
\phi_{L,n}(\ln \bY \mid {\bf{1}}_L \otimes \bmu', {\bf{I}}_L, \bSigma) \\
&\times &\Phi_{Lm} ({\bf{I}}_L \otimes\bDelta'\bSigma^{-1/2} vec(\ln \bY) \mid {\bf{1}}_L \otimes(\bDelta'\bSigma^{-1/2}\bmu),
{\bf{V}}^* ), \nonumber
\end{eqnarray}
where  ${\bf{V}}^*= {\bf{I}}_{Lm} - {\bf{I}}_L \otimes \bDelta'\bDelta$  and $\mathbf{A} \otimes \mathbf{B}$ denotes the Kronecker product of
$\mathbf{A}$ and $\mathbf{B}$, $vec(\cdot)$ is the operator vec and $\phi_{L,n}(\cdot \mid  M ; C, V)$
denotes the pdf of a matrix-variate normal distribution where  $M$ is
an $L\times n$ constant vector and $C$ and $V$ are, respectively, $L \times L$
and $n \times n$ constant matrices. Observe that the likelihood
function in (\ref{EqVeroG1}) defines a class of matrix-variate log-CFUSN
distributions.

In this work, inference is done under the Bayesian paradigm. Therefore we need to specify prior
distributions for all parameters.  We consider $m$ as a fixed constant and also assume
some usual prior distributions for the location   and scale  parameters. In the following
proposition, we present the posterior full conditional distributions for $\bmu$, $\bSigma$ and $\bDelta$ whenever
the prior distributions for $\bmu$, $\bSigma$
are, respectively,
\begin{eqnarray}
\left.\begin{array}{rcl}
\bmu &\sim& N_n (\bmu_{0}, \bSigma_{\mu}) \\
\bSigma &\sim& IW(d, D),
\end{array}\right\}
\label{EqPriorG}
\end{eqnarray}
where $\bmu_{0} \in \RR^n$, $\bSigma_{\mu}$ is an $n \times n$ symmetric, positive definite matrix,
$D$ is an $n \times n$  constant matrix,  $d \in \RR_+$ with $d >n$, and $IW(d, D)$
denotes the inverse-Wishart distribution with parameters $d$ and $D$. A flat prior
distribution for $\bSigma$ is obtained by setting $d$ close to zero.

\begin{prop}
Let $\bY_1, \dots, \bY_L \mid \bmu, \bSigma, \bDelta \buildrel iid \over\sim LCFUSN_{n,m}(\bmu, \bSigma, \bDelta)$.
Assume that, {\it a priori}, the parameters $\bmu$, $\bSigma$ and $\bDelta$ are independent and
that the prior distributions for  $\bmu$, $\bSigma$  are given in (\ref{EqPriorG}). Suppose 
$\bDelta$ has a proper prior distribution $\pi(\bDelta)$. Then, the posterior full conditional distributions for
$\bmu$, $\bSigma$ and $\bDelta$  are given, respectively, by
\begin{eqnarray*}
\pi(\bmu \mid \bSigma, \bDelta, \bY) &\propto&  \phi_n(\bmu \mid \bSigma^*
[\bSigma_{\mu}^{-1} \bmu_{0} + (\bSigma^{-1}\otimes {\bf{1}}_L)' vec(\ln \bY)] \mid \bSigma^* )     \\
&\times & \Phi_{mL}({\bf{I}}_L \otimes[{\bf{I}}_m - \bDelta' \bDelta]^{-1/2}\bDelta'\bSigma^{-1/2}[vec(\ln \bY) - {\bf{1}}_L \otimes \bmu])\\[2mm]
\pi(\bSigma \mid \bmu,\bDelta, \bY ) &\propto& {\mathcal{I}}{\mathcal{W}}_n(d+L+1, D + [\ln\bY - {\bf{1}}_L \otimes \bmu']'[\ln\bY - {\bf{1}}_L \otimes \bmu'])\\
& \times& \Phi_{mL}({\bf{I}}_L \otimes[{\bf{I}}_m - \bDelta' \bDelta]^{-1/2}\bDelta'\bSigma^{-1/2}[vec(\ln \bY) - {\bf{1}}_L \otimes \bmu])\\[2mm]
\pi(\bDelta \mid \bmu, \bSigma, \bY) &\propto& \pi(\bDelta)\Phi_{mL}({\bf{I}}_L \otimes[{\bf{I}}_m - \bDelta' \bDelta]^{-1/2}\bDelta'\bSigma^{-1/2}[vec(\ln \bY) - {\bf{1}}_L \otimes \bmu])\\
\end{eqnarray*}
where $\bSigma^* = [L \bSigma^{-1} + \bSigma_{\mu}^{-1}]^{-1}$ and ${\mathcal I}{\mathcal W}_n(a, A)$ denotes
the pdf of the inverse-Wishart distribution with parameters $a$ and $A$.
\label{ProFCDG}
\end{prop}

The proof of Proposition \ref{ProFCDG} follows  by mixing (\ref{EqVeroG1}), (\ref{EqPriorG}) and $\pi(\bDelta)$
using the Bayes's theorem and some well-known results of matrix theory.  It is noteworthy that the
posterior full conditional distribution of $\bmu$ belongs to the SUN family of distributions.

The univariate case is presented in the following corollary.
Denote by $IG(\alpha, \beta)$,  $\alpha>0$ and $\beta>0$, the inverse-gamma distribution with
$E(\sigma^{2}) = \alpha(\beta-2)^{-1}$.

\begin{cor}
Let $Y_{1},..., Y_{L}\mid \mu, \sigma, \bDelta \buildrel iid \over\sim LCFUSN_{1,m}(\mu,\sigma^2,\bDelta)$
and assume that, a priori, $\mu$, $\sigma$ and $\bDelta$ are independent and such that $\mu \sim N(\mu_0,v)$, 
$\sigma^{2} \sim IG(\alpha, \beta)$, where $l \in \RR$,  $v$, $\alpha$ and $\beta$
are non-negative numbers, and $\bDelta$ has a proper prior distribution $\pi(\bDelta)$. Then,
the posterior full conditional distributions for $\mu$, $\sigma$ and $\bDelta$  are given, respectively, by 
\begin{eqnarray*}
  f(\mu \mid \by, \sigma^{2}, \bDelta)
  &\propto&\phi\left(\mu \Big| \frac{v^{2} {\bf{1}}_L' \ln \by +\mu_0\sigma^{2}}{Lv^{2}+  \sigma^{2}}, \frac{v^{2}\sigma^{2}}{Lv^{2}+\sigma^{2}}\right) \\
  & \times&\Phi_{mL}(\sigma^{-1}(\mathbf{I}_{L} \otimes \bDelta') (\ln \by- \mu {\bf{1}}_L)\mid \mathbf{I}_{mL}- \mathbf{I}_{L}
  \otimes\bDelta'\bDelta)\\ [2mm] 
  f(\sigma^{2} \mid \bx, \mu, \bDelta) &\propto& \left(\frac{1}{\sigma^{2}}\right)^{\frac{L+2\alpha+2}{2}}
  \exp{\left( \frac{2\beta- (\ln \by - \mu {\bf{1}}_L)' (\ln \by - \mu {\bf{1}}_L)}{2\sigma^{2}}\right)}\\ \nonumber
  & \times& \Phi_{mL}(\sigma^{-1}(\mathbf{I}_{L} \otimes \bDelta') (\ln \by- \mu {\bf{1}}_L)\mid \mathbf{I}_{mL}- \mathbf{I}_{L} \otimes\bDelta'\bDelta)\\ [2mm]
  f(\bDelta \mid \bx, \mu, \sigma^{2}) &\propto& \pi(\bDelta)\Phi_{mL}(\sigma^{-1}(\mathbf{I}_{L} \otimes \bDelta') (\ln \by- \mu {\bf{1}}_L)\mid \mathbf{I}_{mL}- \mathbf{I}_{L} \otimes\bDelta'\bDelta),
\end{eqnarray*}
where $\ln \by = (\ln y_1, \dots, \ln y_L)'$.
\label{CoBA1}
\end{cor}

This result is a straightforward consequence of Proposition \ref{ProFCDG}. It follows by observing that
the likelihood function of $\by$ is given by
\begin{eqnarray}
\label{vero1lcfsun}
f(\by | \mu, \sigma^{2}, \bDelta)
&=& 2^{Lm} (2\pi\sigma^{2})^{-L/2}\left(\prod_{i=1}^{L}y_{i}\right)^{-1} \exp\left\{-\sum_{i=1}^{L}
\frac{(\ln y_{i}-\mu)^{2}}{2\sigma^{2}}\right\}  \\
&\times& \prod_{i=1}^{L} \Phi_{m}(\bDelta'\sigma^{-1}(\ln y_{i}-\mu)|\mathbf{I}_{m}-\bDelta'\bDelta),\nonumber 
\end{eqnarray}
and that the inverse-Wishart distribution is a generalization of the multivariate inverse-gamma distribution.

Since the parameter $\bDelta$ is an $ n \times m$ vector with $||\bDelta \textbf{a}||<1$, for all unitary vectors $\textbf{a}
\in \mathbb{R}^{m}$, the elicitation of a prior distribution for $\bDelta$ becomes a hard task.
To overcome this difficulty, we can assume  an alternative parametrization of the model by setting
$\bDelta= \bLambda(\mathbf{I}_{m}+\bLambda'\bLambda)^{-1/2}$ for some $n \times m$ real matrix $\bLambda$.
A possible prior distribution for $\bLambda$ is a multivariate normal distribution. The calculation of the
full conditional distributions under these choices is similar to that
presented in Proposition  \ref{ProFCDG} and thus will be omitted. However, we remark that the posterior
full conditional distributions  for $\bmu$ and $\bLambda$ belong to the SUN class  of distributions
and a skewed inverse-Wishart distribution is the posterior full conditional distribution for $\bSigma$.
Consequently, by considering this class of joint prior distributions for $(\bmu,  \bSigma, \bLambda)$
we have conjugacy. It is notable that we are also performing a conjugate analysis for 
$(\bmu,  \bSigma)$ in the cases discussed in  Proposition \ref{ProFCDG} and Corollary  \ref{CoBA1}.

Another way to overcome the problem is  to assume $\bDelta = \delta {\bf{1}}_{n,m}$
where $\delta$ is a real number belonging to the interval $(-1, 1)$. By carrying this out,
the model loses  some flexibility. On the other hand we obtain a more parsimonious model which is still able to
accommodate different degrees of asymmetry. From now on, we consider this  approach and elicit
a non-informative uniform prior distribution for $\delta$.  Under this more parsimonious model, 
the  posterior full conditional distributions for all parameters follow from Proposition \ref{ProFCDG}
and are given by
 \begin{eqnarray*}
\pi(\bmu \mid \bSigma, \bDelta, \bY) &\propto&  \phi_n(\bmu \mid \bSigma^*
[\bSigma_{\mu}^{-1} \bmu_{0} + (\bSigma^{-1}\otimes {\bf{1}}_L)' vec(\ln \bY)] \mid \bSigma^* )     \\
&\times & \Phi_{mL}(I_L \otimes[I_m - \delta^2 {\bf{1}}_{m,m}]^{-1/2} \delta{\bf{1}}_{m,n}\bSigma^{-1/2}[vec(\ln \bY) - {\bf{1}}_L \otimes \bmu]),\\[2mm]
\pi(\bSigma \mid \bmu,\bDelta, \bY ) &\propto& {\mathcal I}{\mathcal W}_n(d+L+1, D + [\ln\bY - {\bf{1}}_L \otimes \bmu']'[\ln\bY - {\bf{1}}_L \otimes \bmu'])\\
& \times& \Phi_{mL}(I_L \otimes[I_m - \delta^2 {\bf{1}}_{m,m}]^{-1/2} \delta{\bf{1}}_{m,n}\bSigma^{-1/2}[vec(\ln \bY) - {\bf{1}}_L \otimes \bmu]),\\[2mm]
\pi(\bDelta \mid \bmu, \bSigma, \bY) &\propto& \Phi_{mL}(I_L \otimes[I_m - \delta^2 {\bf{1}}_{m,m}]^{-1/2} \delta{\bf{1}}_{m,n}\bSigma^{-1/2}[vec(\ln \bY) - {\bf{1}}_L \otimes \bmu]).\\
\end{eqnarray*}

A difficulty encountered in inference under this family of distributions
is that, independently of the  model we assume (a general $\bDelta$, $\bDelta = \delta {\bf{1}}_{n,m}$
or the reparametrization $\bLambda$), the skewing function for all posterior full conditional distributions
is the cdf of some $mL$-variate normal distribution. Hence the computational cost for sampling
of the posterior distributions tends to become very high.

\subsection{Data augmentation: Simplifying the computation using the Stochastic representation }

A strategy that greatly facilitates Bayesian inference under complex models is the data augmentation technique.
It consists of including latent variables or unobserved data into the model in order to simplify the computational
procedures \citep{DyMe01}. In the proposed model, we accomplish this by considering the  stochastic representations for the CFUSN family
of distributions obtained by \cite{ArGe05}.

By applying a logarithmic transformation to the data, we can estimate the parameters of
the log-CFUSN distribution via  the CFUSN distribution. Formally, if we consider
the marginal stochastic representation in (\ref{EqSR}), the model in
(\ref{EqVeroG1}) can be hierarchically  represented as follows. Let $\bY_{i} \sim LCFUSN_{n,m}(\bmu,\bSigma,\bDelta)$
and $\bZ_{i}=\ln \bY_{i} \sim CFUSN_{n,m}(\bmu,\bSigma,\bDelta)$. Assume also that $\bDelta = \delta {\bf{1}}_{n,m}$,
$\delta \in (-1, 1)$. Then, it follows that 
\begin{equation}\label{eq1}
\bZ_{i}\buildrel d \over = \delta \bSigma^{1/2}{\bf{1}}_{n,m}|\bX_{i}|+[\bSigma(\bI_{n}-\delta^2{\bf{1}}_{n,n})]^{1/2}\bV_{i} +\bmu,
\end{equation}
where $\bX_{i} \sim N_{m}(\bf{0}, \bI_{m})$, $\bV_{i} \sim N_n(\bf{0}, \bI_{n})$, $\bX_{i}$ and $\bV_{i}$
are independent random vectors and $|\bX_{i}|=(|X_{i1}|,...,|X_{im}|)'$.
As a consequence, the model in (\ref{EqVeroG1}) is equivalent to
\begin{eqnarray}\label{hier}
 \bY_{i}  &=& \exp \bZ_{i} \nonumber\\
  \bZ_{i}|\bX_{i}=\bx_{i}&\sim& N_n( \bmu+ \delta \bSigma^{1/2}{\bf{1}}_{n,m} |\bX_{i}|,
  \bSigma({\bf{I}}-\delta^2{\bf{1}}_{n,n})) \nonumber\\
  \bX_{i} &\sim& N_{m}(\bf{0}, \bI_{m}),
\end{eqnarray}                  
where $\bX_{i}$ is a latent (unobserved) random variable.  This hierarchical  representation of the model
is known as data augmentation strategy  and great facilitates the process of sampling
from the posterior  distributions.

Let $\bZ= (\bZ_1, \dots, \bZ_L)'$ and $|\bX|= (|\bX_1|, \dots, |\bX_L|)'$. Under this hierarchical representation, the likelihood
for the augmented data becomes
\begin{equation}
f(\bZ \mid  \bmu, \bSigma, \delta, \bX) = \phi_{L,n}(\bZ \mid {\bf{1}}_L \otimes \bmu' + \delta \bSigma^{-1/2} {\bf{1}}_{n,m}|\bX|',
{\bf{I}}_L, \bSigma({\bf{I}}-\delta^2{\bf{1}}_{n,n})).
\end{equation}

Assume the prior distributions for $\bmu$ and $\bSigma$ given in (\ref{EqPriorG})
and suppose that, {\it a priori}, $\delta \sim {\mathcal{U}}(-1,1)$. It follows that
the full conditional distributions for the parameter $\bmu$, $\bSigma$ and $\delta$
and for the latent variables $\bX_i$, $i=1, \dots, L$ are, respectively,
\begin{eqnarray*}
\bmu \mid  \bSigma, \delta, \bZ, \bX &\sim& N_n(\bSigma_{\mu}^{*-1} [\bSigma_{\mu}^{-1}\bmu_0 + (\bSigma W_{\delta})^{-1} (\bZ' {\bf{1}}_L - \delta \bSigma^{-1/2}{\bf{1}}_{n,m} |\bX|'{\bf{1}}_L )], \bSigma_{\mu}^{*}), \\
f(\bSigma \mid  \bmu, \delta, \bZ, \bX) &\propto& |\bSigma|^{-L/2} \exp \left\{\frac{-tr[(W_{\delta}\bSigma)^{-1}(\bZ- \bmu^* )' (\bZ- \bmu^* )]}{2} \right\},\\
f(\delta \mid \bmu, \bSigma, \bZ, \bX) &\propto& |W_{\delta}|^{-L/2} \exp \left\{\frac{-tr[(W_{\delta}\bSigma)^{-1}(\bZ- \bmu^* )' (\bZ- \bmu^* )]}{2} \right\},\\
f(\bX_i \mid \bmu, \bSigma, \bZ, \delta\bX_{(-i)} ) &\propto&
\exp \left\{ -\frac{1}{2}\left[ |\bX_i|'[ {\bf{I}}_m + \delta^2 {\bf{1}}_{m,n} \bSigma^{1/2} W_{\delta}^{-1}\bSigma^{-1}\bSigma^{1/2}{\bf{1}}_{n,m}]|\bX_i|\right]\right\}\nonumber \\
&\times& \exp\left\{-\frac{1}{2}\left[- \delta |\bX_i|' {\bf{1}}_{m,n} \bSigma^{1/2} W_{\delta}^{-1} \bSigma^{-1}(\bZ_i -\bmu) 
\right]\right\}\nonumber \\
&\times& \exp\left\{-\frac{1}{2}\left[ - \delta(\bZ_i -\bmu)'W_{\delta}^{-1}\bSigma^{-1}\bSigma^{1/2}{\bf{1}}_{n,m}|\bX_i|\right]\right\},
\end{eqnarray*}
where $\bSigma_{\mu}^{*}= [\bSigma_{\mu}^{-1} + L [\bSigma W_{\delta}]^{-1}]^{-1}$,  $W_{\delta}={\bf{I}}_n - \delta^2{\bf{1}}_{n,n}$
and $\bmu^* = {\bf{1}}_L \otimes \bmu' + \delta \bSigma^{-1/2} {\bf{1}}_{n,m}|\bX|'$.

Notice that by using the stochastic representation, the Gibbs sampler  can be used to sample
from the posterior full conditional distribution of $\bmu$.
The posterior full conditional distributions of $\bSigma$, $\delta$ and $ \bX_i$, $i=1, \dots, n$, have
no closed forms and thus the Metropolis-Hastings algorithm can be used.
Moreover, the hierarchical representation in
$(\ref{hier})$ also allows us to use the software Winbugs to obtain samples from the posterior distributions.
We consider it to analyse the dataset in next section.

\section{Case Study}
\label{SecCS} 

In this section we  analyze  the USA monthly precipitation data recorded from 1895 to 2007.
This dataset is available at the National Climatic Data Center (NCDC) and consists of 1.344 observations
of the US precipitation index (PCL). Denote by $Y_i$ the precipitation index in the $i$th month.

\begin{figure}[h!]                                                                                                                                                   
\centering                                                                                                                                                         
\includegraphics[width=12cm]{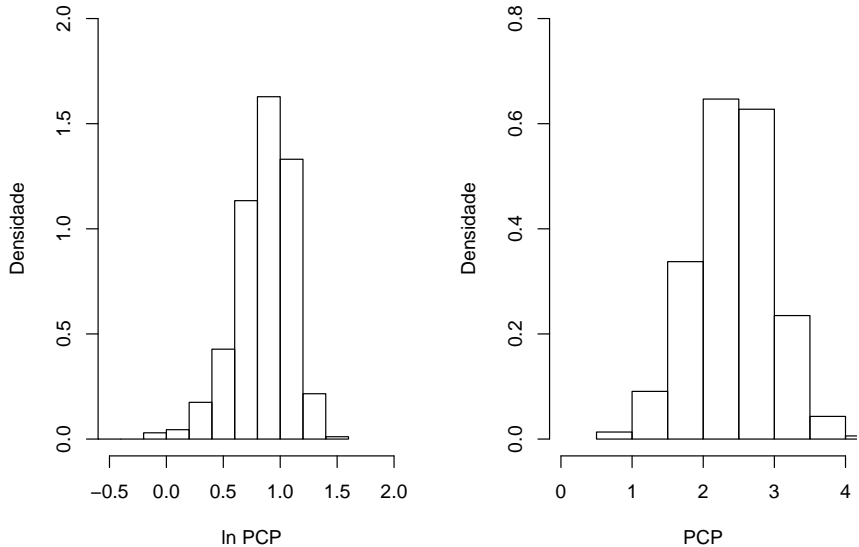}\\
\caption{Histogram of logarithm of PCP (left) and PCP (right).} \label{histN}
\end{figure}

In order to consider the strategy for data analysis described in Section \ref{Sec3},
we consider the log-transformed data.
Figure $\ref{histN}$ shows the histogram for the transformed data (left) and the original data (right),
both of them suggesting the existence of asymmetry in the data, disclosing that
the use of asymmetric distributions can be a reasonable choice to analyze it.

Similar data was previously analyzed by \cite{MaGe10} using the log-skew-normal and the log-skew-$t$
distributions. If compared to the log normal distribution, these models provide a better fit to data.
 \cite{MaGe10} concluded that, due to its flexibility, the log-skew-$t$ distribution, although less parsimonious, worked better than the log-skew normal distribution in capturing the skewness
and heavier tails in the data. 

Depending on $m$, the log-CFSUN family of distributions can be heavier tailed than the log-skew-normal
distribution defined by \cite{MaGe10}. The main goal here is to fit models in the log-CFSUN family of distributions
and evaluate if there is some gain in assuming a higher dimensional skewing function. We consider 
$Y_i \mid \mu, \sigma^2, \bDelta \sim LCFUSN_{1,m}(\mu, \sigma^2, \bDelta)$ and assume the
more parsimonious log-CFSUN family discussed in the previous section where $\bDelta = \delta{\bf{1}}_{m,1}$.
To complete the model specification we assume  flat prior distributions for all parameters
setting  $\mu \sim N(0,100)$, $\sigma^2 \sim IG(0.1, 0.1)$ and $\delta \sim U(-1,1)$.  We
provide a sensitivity analysis considering different values for $m$ ($m=1$ to $5$), which  is assumed to be fixed.
We name $M_i$ the model for which we assume $m=i$.

\begin{figure}[h]
\begin{center}
{\subfigure[$M_1$]{
   \includegraphics[height=3.5cm, width=3.5cm, angle=0]{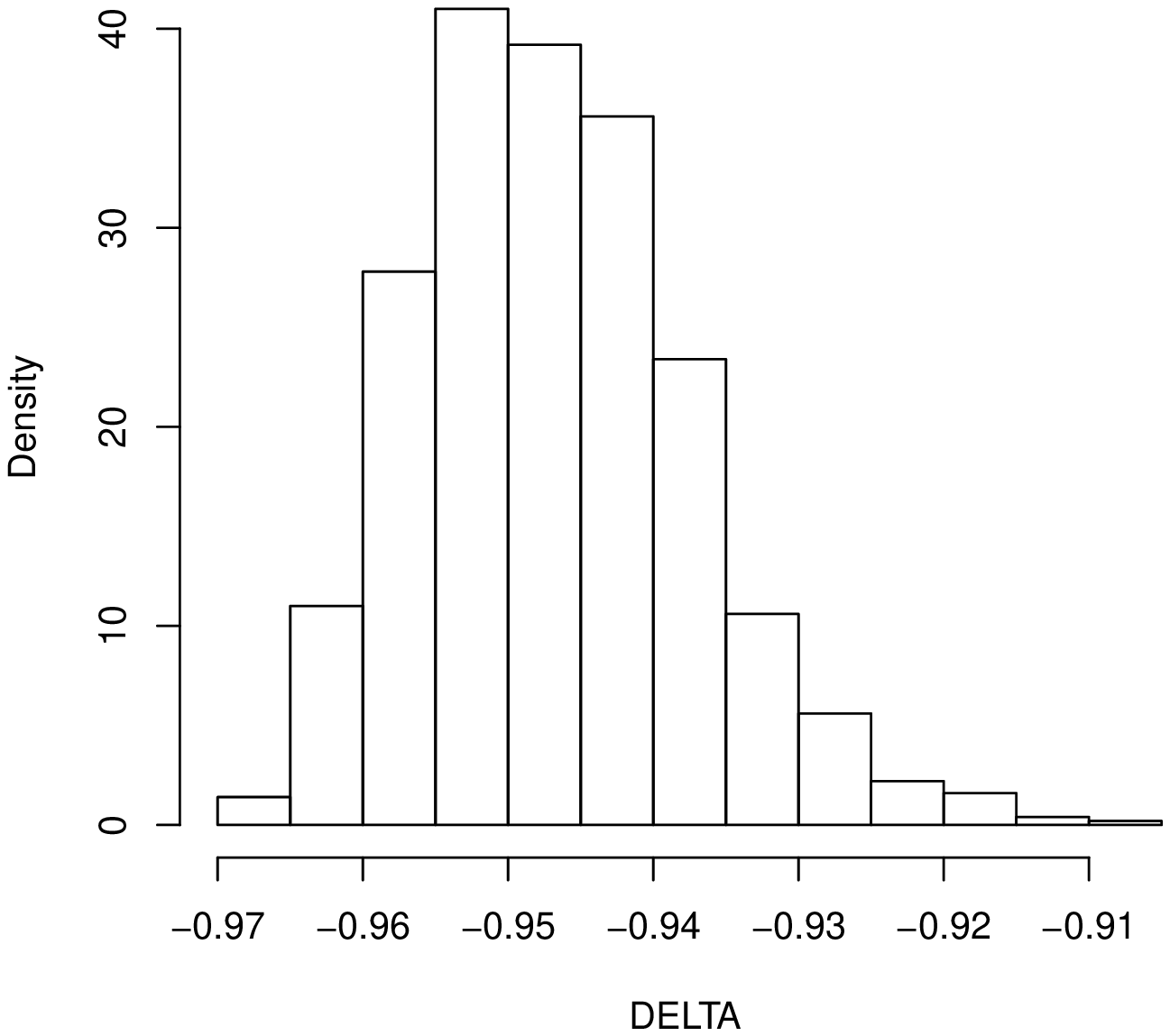}
   \label{d1} }
\subfigure[$M_2$]{
    \includegraphics[height=3.5cm, width=3.5cm, angle=0]{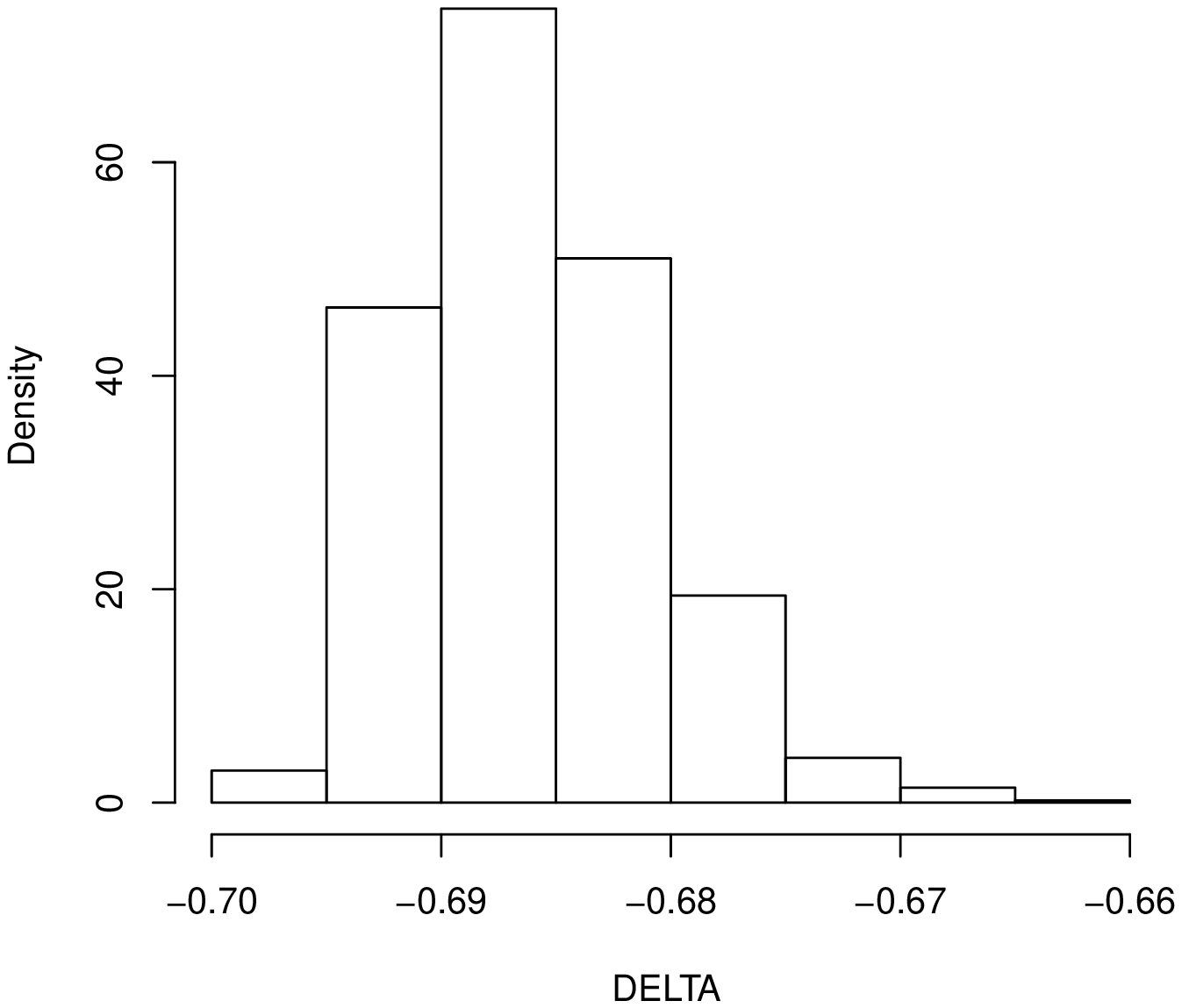}
    \label{d2} }
    \subfigure[$M_3$]{
     \includegraphics[height=3.5cm, width=3.5cm, angle=0]{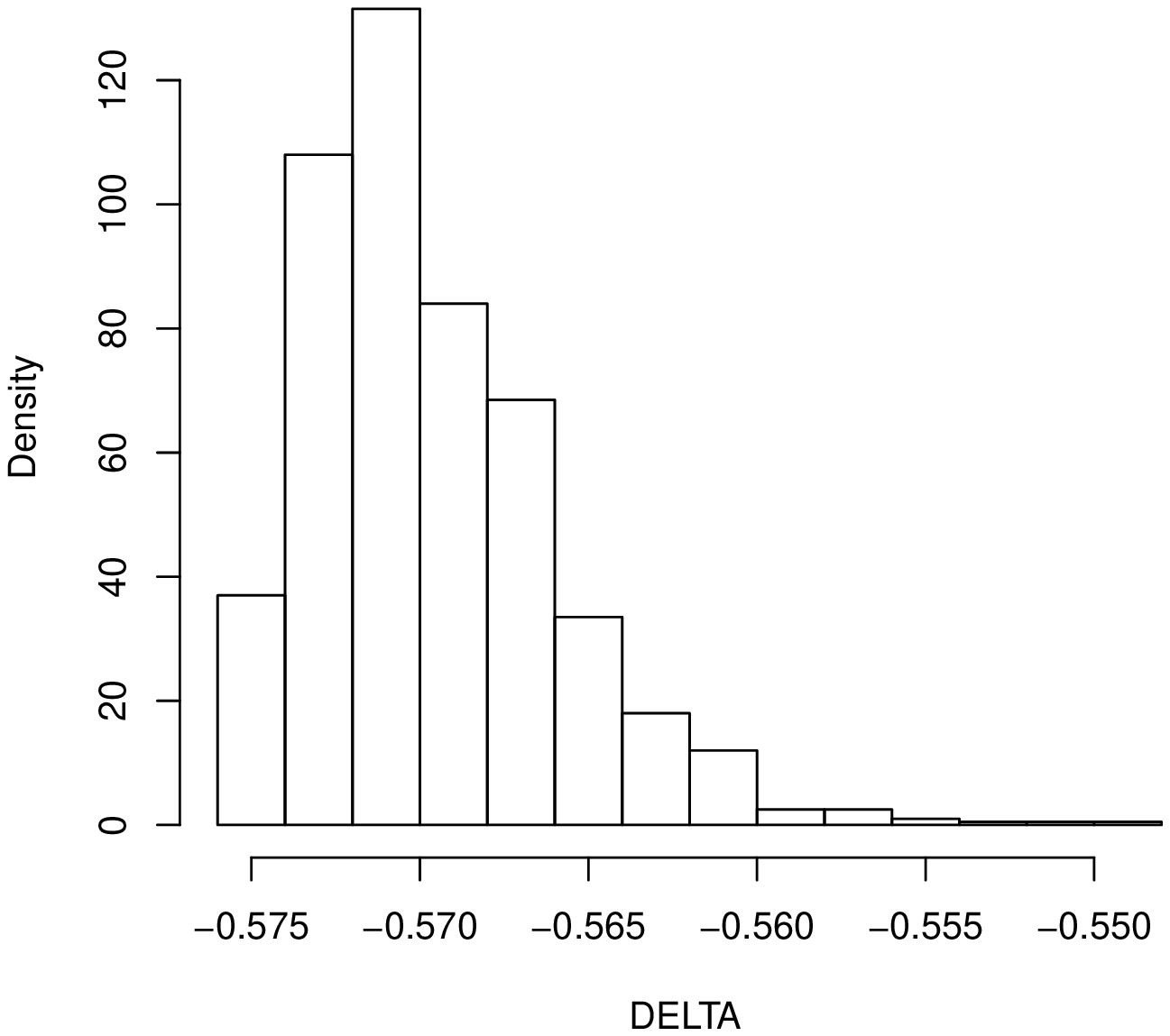}
    \label{d3} }  \\
    \subfigure[$M_4$]{
   \includegraphics[height=3.5cm, width=3.5cm, angle=0]{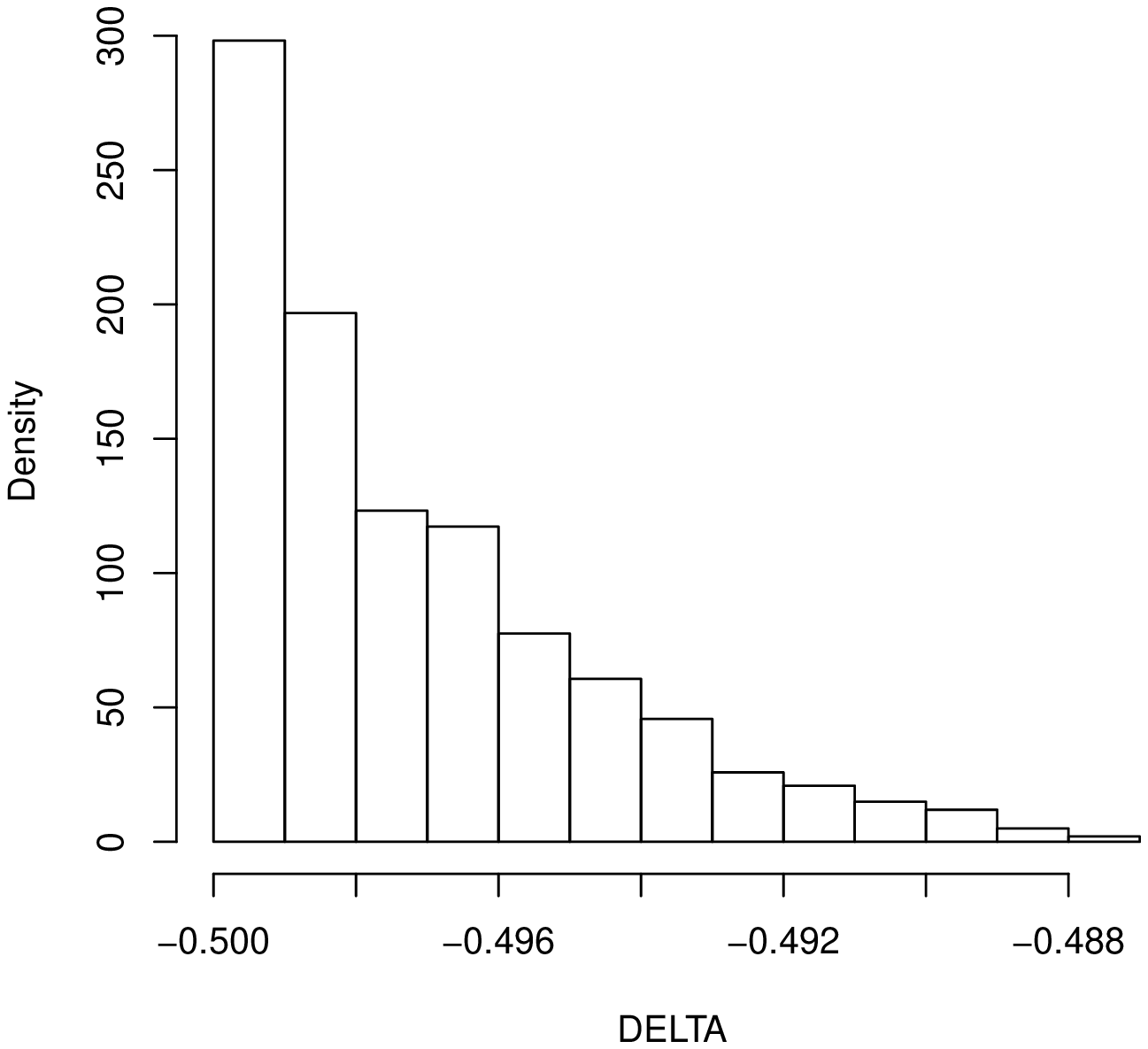}
   \label{d4} }
\subfigure[$M_5$]{
    \includegraphics[height=3.5cm, width=3.5cm, angle=0]{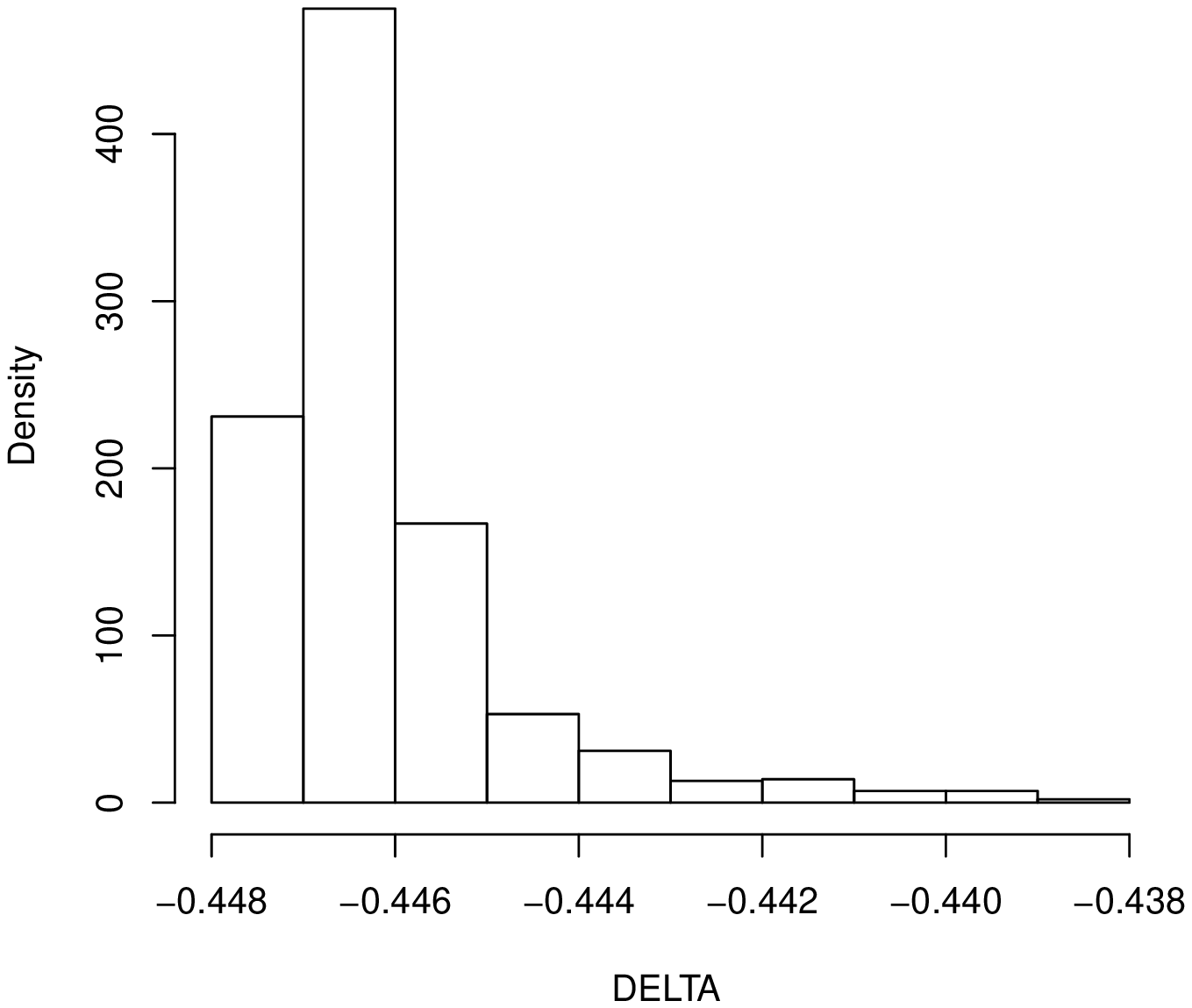}
    \label{d5} }
\caption{\small {Posterior distribution of $\delta$ for all models, precipitation data. }}
\label{PoDel}
}
\end{center}
\end{figure}

Table \ref{TaPoSu} shows some summaries of the posterior distributions
of all parameters. The posterior means for $\mu$ and $\sigma^2$ are similar
for all models and increase as $m$ increases.
Also, all models point out a negative skewness in the data
and the highest estimate for $\delta$ is obtained
if $m=1$, that is, whenever a less dimensional skewing
function is assumed.
It is also noteworthy that the posterior inference about $\mu$ 
is less precise for models with high $m$ since the posterior variance
for that parameter becomes higher as  $m$ increases. The opposite is observed
for $\sigma^2$ and $\delta$.  The $95\%$ HPD intervals disclose  strong evidence
in favour of an asymmetric model with negative skewness (see also Figure \ref{PoDel}
that shows the posterior distribution for $\delta$ in all cases).
\begin{table}[htb]
\begin{center}
\footnotesize
\caption{ Posterior summaries, Precipitation data }
\label{TaPoSu}
\tabcolsep=3pt 
\begin{tabular}{ccccccccccc}
\hline
\multicolumn{1}{c}{}&& \multicolumn{3}{c}{$\mu$} &\multicolumn{3}{c}{$\sigma$} &\multicolumn{3}{c}{$\delta$} \\
\cline{3-4}\cline{6-7}  \cline{9-11}
$m$ & &Mean & St. Dev. & & Mean & St. Dev. & & Mean & St. Dev. & $95\%$HPD  \\ \hline
1 &&  $1.140$ & $0.010$ &&  $0.375$ &  $0.011$ && $-0.947$ & $0.010$ & $[-0.962, -0.925]$\\
2 &&  $1.276$ & $0.013$ &&  $0.384$ &  $0.010$ && $-0.686$ & $0.005$ & $[-0.694, -0.674]$\\
3 &&  $1.392$ & $0.016$ &&  $0.392$ &  $0.010$ && $-0.570$ & $0.004$ & $[-0.575, -0.561]$\\
4 &&  $1.483$ & $0.015$ &&  $0.394$ &  $0.008$ && $-0.497$ & $0.003$ & $[-0.499, -0.490]$\\
5 &&  $1.562$ & $0.015$ &&  $0.394$ &  $0.008$ && $-0.446$ & $0.001$ & $[-0.447, -0.441]$\\
\hline
\end{tabular}
\end{center}
\end{table}

\begin{figure}[h!]
\centering
\includegraphics[width=7.5cm]{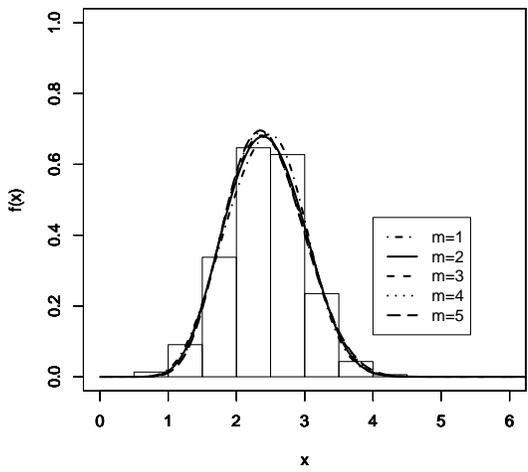}
\caption{Fitted log-CFUSN densities, precipitation data.}
\label{AplicPre}
\end{figure}

Figure \ref{AplicPre} presents the plug-in estimates
of the true density for all $m$ and Table \ref{TaPre}
presents the posterior predictive probabilities of exceeding
the data average ($2.42$), the maximum ($4.20$) and also the probability of not
exceed the minimum ($0.54$). Both informations disclose that the
models are comparable. Moreover, the
predictive summaries show that the left tail of the
posterior predictive  distribution is lighter
than the right one which is in agreement with
the empirical distribution of the data.

\begin{table}[htb]
\begin{center}
\footnotesize
\caption{Posterior Predictive Probabilities, Precipitation data }
\label{TaPre}
\tabcolsep=3pt 
\begin{tabular}{cccc}
\hline
$m$ & $Prob>2.42$  & $Prob>4.2$  & $Prob<0.54$  \\ 
\hline
1& $0.5068$ & $ 6.4514\times 10^{-4} $ & $3.2422\times 10^{-6} $\\
2& $0.4952$ & $ 5.2508\times 10^{-4} $ & $1.7958\times 10^{-6} $\\
3& $0.4920$ & $ 3.1177\times 10^{-4} $ & $1.3747\times 10^{-6} $ \\
4& $0.4907$ & $ 4.4087\times 10^{-4} $ & $8.4866\times 10^{-7} $ \\
5& $0.4909$ & $ 3.1727\times 10^{-3} $ & $5.7553\times 10^{-7} $ \\
\hline
\end{tabular}
\end{center}
\end{table}

Some measures for  model comparison are presented in Table \ref{TaMoSe}. Specifically, we consider
the sum of the logarithm of the conditional predictive ordinate (SlnCPO) \citep{GeDe94, Ge96} and  the deviance
information criterion (DIC) \citep{Spieg02,Celeux06}.
Both criteria point out the model with high dimensional skewing
function ($M_5$) as the best model. It is also remarkable that the DIC presents a
monotonic behaviour. The Kolmogorov-Smirnov goodness of fit test comparing
the plug-in estimate and the empirical cdf is also shown in Table  \ref{TaMoSe}. The statistic $D_n$ and the $p$-value
are calculated as  in \cite{LiLeHs07}.  The differences between the empirical and the estimated c.d.f are not significant
and,  differently of DIC and the SlnCPO, the $D_n$ indicates model
$M_1$ as the best one.

\begin{table}[htb]
\begin{center}
\footnotesize
\caption{Model selection statistics, Precipitation data }
\label{TaMoSe}
\tabcolsep=3pt 
\begin{tabular}{ccccccc}
\hline
\multicolumn{1}{c}{}&& \multicolumn{3}{c}{Kolmogorov-Smirnov}  \\
\cline{3-4}
$m$ &&   $D_n$   & P-value    && DIC & SlnCPO  \\ \hline
1   &&  $0.02508$ & $0.33978$ && $-13,190$ & $-0.83766$\\
2   &&  $0.02765$ & $0.25874$ && $-36,960$ & $-0.83545$\\
3   &&  $0.03033$ & $0.17621$ && $-112,400$& $-0.83765$\\
4   &&  $0.03244$ & $0.11524$ && $-321,100$& $-0.84144$\\
5   &&  $0.03082$ & $0.16208$ && $-895,300$& $-0.81057$\\
\hline
\end{tabular}
\end{center}
\end{table}

\section{Conclusions}
\label{SecCo} 
 
In this paper we introduced two classes of log-skewed distributions
with normal kernels: the log-CFUSN and the log-SUN. We studied some properties
of the log-CFUSN family of distributions such as  marginal and
conditional distributions, moments and stochastic representation.
We also discussed some issues related to Bayesian inference in that family.
Our discussion was devoted to the elicitation of a prior
distribution for the skewness parameter.

The main motivation for studying the log-CFUSN family of distribution in detail,
and other new classes of log-skewed distributions, is the result
that appeared in \cite{SaLoAr13} where it was shown that
such family is of fundamental interest in the interpretation of the parameters in mixed logistic
regression model if the random effects are skew-normally distributed.
In that paper it was proved that, under skew-normality,  
the odds ratio has distribution in the log-CFUSN family.

Analizing the USA precipitation dataset, we concluded that the
use of a skewing function with  higher dimension than that
assumed by \cite{MaGe10} can bring
some gain to the model fit.

\section*{Acknowledgement}
The authors would like to thank the Editors and the referee for their  comments
and suggestions which improved the paper.
We would like to express our gratitude to Professors Fredy Castellares
and Reinaldo Arellano for their comments on the first draft of this work.
The research of M.M. Queiroz  was partially supported by CAPES (\emph{Coorde\-na\c c\~ao  de Aperfei\c coamento
de Pessoal de N\'{\i}vel Superior}) and CNPq (\emph{Conselho Nacional de Desenvolvimento
Cient\'{\i}fico e Tecnol\'{o}gico}) of the Ministry for Science and Technology of Brazil.
R. H. Loschi would like to thank to CNPq, grants 301393/2013-3 and 306085/2009-7,
for a partial allowance to her researches. The research of R.W.C. Silva was
partially supported by PRPq-UFMG (Edital 12/2011).

\end{document}